\documentclass[%
 reprint,
superscriptaddress,
 amsmath,amssymb,
 aps,
prb
]{revtex4-1}
\usepackage{graphicx,hyperref,amsthm,xcolor,soul,braket,todonotes,amssymb,amsmath,tikz-cd,pst-node,lineno}
\usetikzlibrary{cd}
\usepackage[]{subcaption}
\usepackage{dcolumn}
\usepackage{bm}
\usepackage{natbib}
\pdfoutput=1
\setcitestyle{numbers,square}
\hypersetup{
    linkcolor=blue,
    citecolor=blue,
    filecolor=blue,
    urlcolor=blue,
    colorlinks=true
}
\newenvironment{nalign}{
    \begin{equation}
    \begin{aligned}
}{
    \end{aligned}
    \end{equation}
    \ignorespacesafterend
}

\begin{document}
	\title{Analysis of Many-body Localization Landscapes and Fock Space Morphology via Persistent Homology}
	
	\author{Gregory A. Hamilton }
	\email{gah4@illinois.edu}
	\author{Bryan K. Clark}
	\affiliation{ 
Institute for Condensed Matter Theory and IQUIST and Department of Physics, University of Illinois Urbana-Champaign, Urbana, IL 61801, USA}
	\date{\today}

	\begin{abstract}
We analyze functionals that characterize the distribution of eigenstates in Fock space through a tool derived from algebraic topology: persistent homology. Drawing on recent generalizations of the localization landscape applicable to mid-spectrum eigenstates, we introduce several novel persistent homology observables in the context of many-body localization that exhibit transitional behavior near the critical point. We demonstrate that the persistent homology approach to localization landscapes and, in general, functionals on the Fock space lattice offer insights into the structure of eigenstates unobtainable by traditional means.
    \end{abstract}

\maketitle

\section{Introduction}

Many-body localization (MBL) is a paradigmatic example of a dynamical phase transition whereby eigenstates fail to thermalize \cite{Abanin_Altman_Bloch_Serbyn_2019,Thomson_Schiro,Scardicchio_Thiery_2017}. MBL is characterized by a multitude of observables including vanishing conductivity \cite{alet2018many,gopalakrishnan2015low}, logarithmic entanglement growth \cite{kim2014local}, area law entanglement \cite{Pekker_Clark_2017}, level statistics \cite{Alet_Laflorencie_2018}, quasilocal integrals of motion \cite{Roeck_Imbrie_2017,Imbrie_Ros_Scardicchio_2017,Thomson_Schiro}, and a litany of others \cite{Luitz_Laflorencie_Alet_2015, Abanin_Altman_Bloch_Serbyn_2019}. 
A variety of models feature in the MBL literature, including the disordered model considered here, quasiperiodic potentials \cite{Cookmeyer_Motruk_Moore_2020}, and quadratic Stark potentials \cite{Schulz_Hooley_Moessner_Pollmann_2019}. 
While microscopic mechanisms for the phase transition have been proposed and elucidated upon \cite{Morningstar_Huse_Imbrie_2020,Monteiro_Micklitz_Tezuka_Altland,Vidmar_Krajewski_Bonca_Mierzejewski_2021,roy2019self}, much controversy still surrounds the stability of the MBL phase in the thermodynamic limit \cite{imbrie2016diagonalization,Chandran_Laumann_Oganesyan_2015} as well as the structure of eigenstates near the transition \cite{roy2022hilbert}. \par 
In response to this challenge a large body of recent work probes the correlational structure of eigenstates in Fock space (here taken to be the configurational $\sigma_{z}$ basis). The guiding intuition for this approach is, as the many-body Hamiltonian maps to a tight binding model (TBM) on a Fock space lattice (or graph), eigenstate spatial correlations on the lattice dictate the behavior of autocorrelation functions that diagnose the phase transition \cite{Roy_Logan_2021a}. \par 
Fock space eigenstate structure has been a fundamental question since the early days of Anderson localization (1BL), as the inverse participation ratio (IPR) and participation entropies $S_{i}^{P}$ serve as indicators for the Anderson transition \cite{Luitz_Laflorencie_Alet_2015,Mace_Alet_Laflorencie_2019,Tikhonov_Mirlin_2021,Scardicchio_Thiery_2017}. From the participation entropy stems a notion of multifractality with multifractal exponents identifying the universality class; an analogous approach to the participation entropy exists in the many-body case. \par In pursuit of a more nuanced understanding of the transition and MBL stability, recent approaches look beyond the participation entropies to quantify the ``anatomy" of Fock space correlations \cite{Roy_Chalker_Logan_2019,Roy_Logan_2021a,Roy_Logan_2021b,roy2019self}. From this TBM picture several phenomenological models for the MBL transition have been proposed, including so-called ``avalanche" mechanisms \cite{Tomasi_Khaymovich_Pollmann_Warzel_2020,Szoldra_Sierant_Kottmann_Lewenstein_Zakrzewski_2021,morningstar2022avalanches}, and, pertinent to this work, percolation models \cite{Roy_Chalker_Logan_2019,Roy_Logan_Chalker_2019,Goblot_Strkalj_Pernet_Lado_Dorow_Lemaitre_Gratiet_Harouri_Sagnes_Ravets,Roy_Logan_2021a,Roy_Logan_2021b,Logan_Welsh_2019,Thiery_Huveneers_Muller_De,Prelovsek_Barisic_Mierzejewski_2018}. We approach this Fock space viewpoint from a morphological lens by leveraging a new functional: the localization landscape. \par

New insights into the 1BL problem hinge on a functional called the \textit{localization landscape} (LL) \cite{filoche2012universal}, which bounds remarkably well the spatial extent of low-energy eigenstates \cite{filoche2012universal,Arnold_David_Filoche_Jerison_Mayboroda_2018,Arnold_David_Jerison_Mayboroda_Filoche_2015}. The LL is the solution to a simple differential equation: \begin{align}
    H\bm{u} = \bm{1},
\end{align} for Hamiltonian $H$ that satisfies mild constraints ($\bm{1}$ is the vector of all ones). The LL (or rather, its inverse $V_{u} = 1/\bm{u}$) operates as an ``effective potential", and can be used in place of the physical potential to well approximate both the integrated density of states via the classical Weyl law \cite{Arnold_David_Jerison_Mayboroda_Filoche_2015,Pelletier_Delande_Josse_Aspect_Mayboroda_Arnold_Filoche_2021,Shamailov_Brown_Haase_Hoogerland_2021} and localization lengths for low-energy eigenstates \cite{Shamailov_Brown_Haase_Hoogerland_2021}. The effective potential, being numerically tractable, has already been used extensively in the context of quantum transport in disordered semiconductors \cite{li2017localization,Pelletier_Delande_Josse_Aspect_Mayboroda_Arnold_Filoche_2021,Arnold_David_Jerison_Mayboroda_Filoche_2015}. \par 

Connections between percolation theory and transport extend back several decades \cite{kirkpatrick1973percolation}, and the effective potential $V_{u}$ lends itself to an interpretation wherein a transition from localized to extended eigenstates is percolative, as has been recently hypothesized \cite{li2017localization,Balasubramanian_Liao_Galitski_2020}. This core idea, that the morphological shift in an effective potential (or more generally, a potential in a configuration or phase space) might serve as an indicator of a phase transition, has been well-developed in the context of classical phase transitions and serves as the impetus for our work here \cite{Bel-Hadj-Aissa_Gori_Franzosi_Pettini_2021,Gori_Franzosi_Pettini_Pettini_2022,Sale_Giansiracusa_Lucini_2022,donato2016persistent}.  \par 

As we describe in Sec. \ref{sec:theory}, the LL is perfectly well-defined for the many-body case by the TBM mapping \cite{Balasubramanian_Liao_Galitski_2020}, and variants of the LL equally apply to interior eigenstates \cite{Herviou_Bardarson_2020,lemut2020localization}. The localization landscape variant we use here (discussed in detail in Sec. \ref{sec:theory}) is termed the $L_{2}$ landscape, and takes the form \begin{align}
    (u^{(2)})^{2} = \text{diag} (M^{-1}), \, M := H^{\dagger}H,
\end{align} where $H$ is the Hamiltonian \cite{Herviou_Bardarson_2020}. As we discuss in Sec. \ref{sec:theory}, with a proper choice of energy $E_{0}$ the $L_{2}$ landscape bounds mid-spectrum eigenstates just as the LL bounds low-energy eigenstates. In what follows we analyze the $L_{2}$ landscape of a 1D spinless fermion model with open boundary conditions, given by
\begin{nalign}\label{eq:ham}
    H = H_{t} + H_{W} + H_{V}  \\
 H_{t}  = t\sum_{\langle ij\rangle }\left(c_{i}^{\dagger}c_{j} + \text{ H.c.}\right), \\
 H_{W} = \sum_{i}h_{i}\left(n_{i}-\frac{1}{2}\right), \\  
 H_{V} = V\sum_{\langle ij \rangle}\left(n_{i}-\frac{1}{2}\right) \left(n_{j}-\frac{1}{2}\right), 
\end{nalign} with $P(h_{i}) \in [-W,W]$ uniform and $\langle ij\rangle$ denoting nearest neighbor (NN) pairs. The dimensionless parameter $V/t$ sets the strength of NN interactions; for a choice of $(t,V) = (1/2,1)$ this model reduces to the standard XXZ model. 
The $L_{2}$ landscape, while simple to define, is a complex functional on the Fock space lattice. To ascertain what the $L_{2}$ landscape can tell us about the MBL transition, we leverage a new tool from computational geometry and topological data analysis: persistent homology. \par 
Persistent homology offers a computable handle on the small and large-scale structure of a point cloud or underlying functional by building a \textit{filtration} of topological objects, usually simplicial complexes. Persistent homology is largely parameter-free and yields novel observables fundamentally tied to discrete Morse theory and algebraic topology \cite{Zomorodian_Carlsson_2005}. While persistent homology has recently been used to identify classical and quantum phase transitions \cite{He_Xia_Angelakis_Song_Chen_Leykam_2022,Cole_Loges_Shiu_2021,Olsthoorn_Balatsky_2021}, we leverage it here to gain a deeper understanding of whether the $L_{2}$ landscape exhibits signs of a percolation transition. \par 
To this end we generate a large set of observables that quantify the morphology structure of the LL with respect to the high-dimensional Fock space. The observables and topological summaries we garner from this approach paint a complex picture. While some results are easy to interpret in the context of standard observables, still others defy simple interpretations. Part of this difficulty lies in the incoincidence of persistent homology with standard statistical techniques, an impediment we confront several times in this work. \par 
Nonetheless, we do find indications that persistent homology accesses the fundamental structure of mid-spectrum eigenstates and coincides with critical points found in previous works. We provide novel ways to probe the ``anatomy" of Fock space at the morphological level. In Sec. \ref{sec:conclusion} we describe many more ways future research can extend and refine our approach.

To briefly state the novel contributions of our work: 
\begin{itemize}
    \item We describe a new tool, persistent homology, to probe the morphology of Fock space. In contrast to recent investigations of Fock space correlations \cite{Prelovsek_Barisic_Mierzejewski_2018}, this technique is parameter-free and probes clustering at all scales.
    \item We identify the proliferation of extremal points on the $L_{2}$ as an indicator for the MBL transition. 
    \item We demonstrate several topological observable that probe the clustering properties of the $L_{2}$ landscape scale with the Hilbert space dimension. The scaling exponents exhibit cross-over behavior near the MBL transition. 
    \item We give a new, persistent homology-based definition for fractal dimensions applicable to both many-body eigenstates and functionals on Fock space.
\end{itemize}

The outline of this exploratory work is as follows. In Sec. \ref{sec:theory} we describe the localization landscape, its extension to the many-body case, and the variant on the LL utilized in this work. We also give an introduction to persistent homology and its primary output, the persistence diagram. In Sec. \ref{sec:results} we apply our persistent homology pipeline to the Hamiltonian in Eq. \ref{eq:ham}. We give a phenomenological understanding of several observables extracted from the persistent homology pipeline, and assess the implications for percolative MBL phase transition hypothesis. Finally, in Sec. \ref{sec:conclusion} we summarize our work and suggest future directions of study.



\begin{figure*}
    \centering
    \includegraphics[width=\textwidth,trim={0cm 5cm 0cm 4cm},clip]{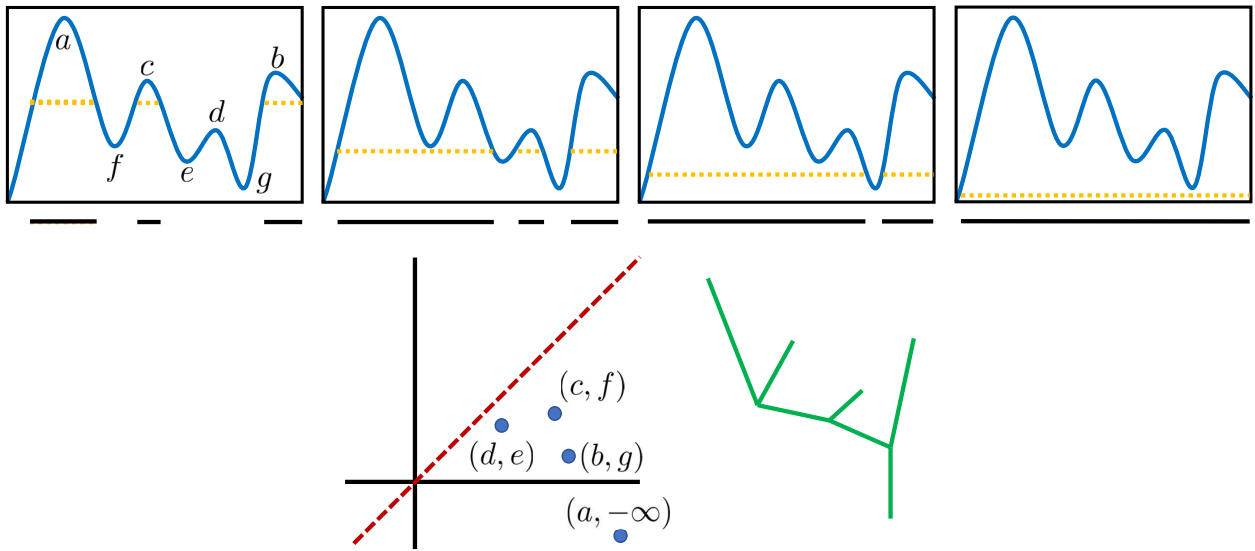}
    \caption{Pictorial representation of a superlevel set filtration for a one-dimensional function. The dashed line denote different level sets, while the black lines underneath each box denote the intervals in the superlevel set. As the level decreases, intervals emerge and merge until a final connected components is formed. The bottom plot shows the corresponding persistence diagram. The rightmost point $(a,-\infty)$ is displayed but is ignored with respect to topological summaries of the persistence diagram. The green dendrogram denotes the merge tree derived from the subset filtration. External vertices correspond to the maxima, while internal vertices correspond to the minima. }
    \label{fig:superlevel}
\end{figure*}

\section{Theory} \label{sec:theory}

In this section we overview the LL and $L_{2}$ landscape. We motivate and describe how to apply persistent homology to Fock space functionals before describing topological summaries of Fock space morphology extracted from persistence diagrams and hierarchical clustering.

\subsection{Localization landscapes}

The original LL construction takes as input a single-body Hamiltonian $H$ with eigenspectrum $\{(E^{\beta},\ket{\phi^{\beta}})\}$, expressed in orthonormal basis $\{\ket{I}\}$ such that $H$ is positive-definite and $H_{IJ} < 0$ for all $I\ne J$. For Eq. \ref{eq:ham} these conditions are easily achieved by adding an offset to $H \mapsto -H + \Gamma I$ to ensure positivity (though care should be taken regarding the choice of $\Gamma$ \cite{Filoche_Mayboroda_Tao_2021}). Then the solution $u$ to the equation \begin{align}\label{eq:ll}
    Hu = 1
\end{align} is component-wise positive. Alternatively, we can express $u$ as \cite{Herviou_Bardarson_2020} \begin{align}\label{eq:u}
    u_{J} = \sum_{\beta} \frac{\phi_{J}^{\beta}}{E^{\beta}}\sum_{I}\phi_{I}^{\beta},
\end{align} where $\phi_{J}^{\beta} = \braket{J|\phi^{\beta}}$. \par 
Remarkably, $u$ bounds eigenstate amplitudes, as we have \begin{align}\label{eq:upper_bound}
    |\phi_{J}^{\beta}| \le |E^{\beta}|||\phi^{\beta}||_{\infty} u_{J},
\end{align} where $||\phi||^{p}:= \left(\sum_{I} |\phi_{I}|^{p}\right)^{1/p} $ denotes the $\ell_{p}$ norm. The inverse $V_{u}:= 1/u$ serves as an effective potential, the valleys of which indicate where low-energy eigenstates tend to localize. The peak ranges, or ``valley network" of $V_{u}$ (the unstable manifold of $u$ in the context of discrete Morse theory) demarcate the domains of eigenstate support \cite{Shamailov_Brown_Haase_Hoogerland_2021,Herviou_Bardarson_2020}. \par 

To apply the LL to the many-body case, we map the Hamiltonian in Eq. \ref{eq:ham} to a TBM on the Fock space lattice/graph $G_{\mathcal{F}} = (V,E)$. Here we closely follow the notation and setup given in Ref. \cite{Logan_Welsh_2019}, in which several properties of the Fock space lattice and on-site energies are described.

The physical lattice has $N$ sites with $N_{e} =\lceil \nu N \rceil$ fermions at filling fraction $\nu$, set to $\lfloor N/2 \rfloor$ in this work. The dimension of the \textit{Fock space graph} is then $|V| = \, ^{N}C_{N_{e}} = \binom{N}{N_{e}}$, with nodes $V$ indexed by $\{\ket{I}\}$, the eigenstates of $H_{0} = H_{W} + H_{V}$. This is in contrast to the \textit{Anderson basis} which diagonalizes $H_{t}+ H_{W}$. Each $I$ can be represented by a tuple of $N$ occupation numbers $n_{i}^{(I)} = 0$ or $1$ for each physical site $i$ such that $\sum_{i}n_{i}^{(I)} = N_{e}$. We note that $N_{\mathcal{H}} \propto \exp(s_{\infty}N)$ where 
\begin{align}
   s_{\infty} = -\left(\nu \ln \nu + (1-\nu)\ln (1-\nu)\right)
\end{align}
is the per-site configurational entropy \cite{Logan_Welsh_2019}. \par 
The edges of $G_{\mathcal{F}}$ correspond to states (nodes) $I,J$ connected by $H_{t}$, i.e., \begin{align}\label{eq:tbm}
    H = \sum_{I}\mathcal{E}_{I}\ket{I}\bra{I} + \sum_{\langle IJ \rangle }t\ket{I}\bra{J}.
\end{align} 
We depict in the bottom right of Fig. \ref{fig:network_filtration} $G_{\mathcal{F}}$ for $N = 8 = 2N_{e}$. \par 
We denote by $N(I)$ the neighborhood of $I$, such that the coordination number $Z_{I} = |N(I)|$. The average coordination number over $G_{\mathcal{F}}$ is given by \cite{Logan_Welsh_2019} \begin{align}
    \overline{Z} = 2(1-\nu) N_{e}.
\end{align} 
The competition of the extensive coordination number $\overline{Z}$ (favoring delocalization), with the strong correlations of neighboring on-site energies (discussed in Sec. \ref{sec:results}) is fundamental to the enigma of whether MBL survives in the thermodynamic limit \cite{Logan_Welsh_2019}.\par 

Turning back to the LL, the TBM in Eq. \ref{eq:tbm} $u_{2}$ is defined so long as $H \ge 0$, which is always achievable by adding a positive offset. In fact, any \textit{stoquastic} Hamiltonian yields a positive $u$ \cite{Filoche_Mayboroda_Tao_2021}. From $V_{u}$ a derived, degenerate \textit{Agmon} metric on Fock space bounds the extent of low-energy eigenstate spread into classically disallowed regions \cite{Filoche_Mayboroda_Tao_2021}. \par  However, one significant shortcoming of the LL is its ineffectiveness in bounding eigenstates other those close to the ground state. This follows from the definition of $u$ in Eq. \ref{eq:u}: low-energy eigenstates contribute more to the LL. An analogue dual formulation of the LL can be made for eigenstates at the top of the spectrum \cite{lyra2015dual}, but both parts of the spectrum localize at zero disorder strength in the thermodynamic limit for the Hamiltonian in Eq. \ref{eq:tbm}. Thus, the traditional LL is not well-equipped to probe mid-spectrum eigenstate Fock space anatomy.\par 

Directly in response to this limitation, a new localization landscape was introduced in Ref. \cite{Herviou_Bardarson_2020}, termed the $L_{2}$ landscape. The function $u$ is modified to take the form \begin{align}\label{eq:u2}
    u_{I}^{(2)} := \sqrt{M^{-1}_{II}}, \, M = H^{\dagger}H.
\end{align} Similar to Eq. \ref{eq:upper_bound}, this new landscape satisfies \begin{align}
    |\phi_{I}^{\beta}|\le |E^{\beta}|||\phi^{\beta}||_{2}\sqrt{M^{-1}_{II}},
\end{align} where by orthonormality the $\ell_{2}$ norm equals unity (hence the name ``$L_{2}$ landscape"). Since $M>0$ for any $H$, we shift $H \to H - (E_{0}+i\eta)$  and probe the eigenspectrum near $E_{0}$, where $\eta \in \mathbb{R}^{+}$ is an infinitesimal offset. From this shift the $L_{2}$ landscape can be written as a function of the resolvent $G^{\pm}(E_{0}):=(E_{0}\pm i\eta -H)^{-1}$: \begin{align}
    (u^{(2)}_{J}(E_{0}))^{2} = \left(G^{+}(E_{0})G^{-}(E_{0})\right)_{JJ},
\end{align}
which for finite systems $u^{(2)}$ is exactly given by \begin{align}\label{eq:u2_easy}
    (u_{J}^{(2)}(E_{0}))^{2}  = \sum_{\beta}\frac{|\phi_{J}^{\beta}|^{2}}{|E_{0}+i\eta - E_{\beta}|^{2}}.
\end{align}

Closely related is the local density of states \begin{align}
    \rho_{J}(E_{0}) := \lim_{\eta \to 0^{+}}\frac{1}{2\pi i}\left( G^{+}_{JJ}(E_{0}) - G^{-}_{JJ}(E_{0})\right).
\end{align}
It therefore follows \cite{Herviou_Bardarson_2020} \begin{align}\label{eq:urho_conn}
   \rho_{J}(E_{0}) = \lim_{\eta \to 0^{+}} \eta (u_{J}^{(2)})^{2}.
\end{align}
The behavior of the resolvent $G^{\pm}$ is the central consideration of several recent works \cite{Monteiro_Micklitz_Tezuka_Altland}, which we discuss further in Sec. \ref{sec:results}.\par

The $L_{2}$ landscape incorporates information from the full eigenspectrum, and has a nonlinear dependence on the offset $\eta$, the consequences of which we address in Sec. \ref{sec:results}. Due to the connectivity of $G_{\mathcal{F}}$, it is not immediately obvious what observables can be extracted from the $L_{2}$ landscape. However, the authors in Ref. \cite{Herviou_Bardarson_2020} where the $L_{2}$ landscape was first defined posed an open question of how fractal dimensions like those extracted from the participation entropies can be generalized to the $L_{2}$ landscape. An analogue of the fractal dimension for $\rho(E_{0})$ was recently proposed \cite{Murphy_Wortis_Atkinson_2011}, but we explore in Sec. \ref{sec:results} a novel construction by leveraging a tool originating in Morse theory and rooted in algebraic topology: persistent homology.

\subsection{Persistent Homology: motivation and theory}
Persistent homology characterizes the small and large-scale topological structure of an underlying dataset by computing how the homology of a filtration of topological spaces changes as a function of scale. This fundamental idea underlies the many generalizations of persistent homology; for the interested reader there exist many reviews such as Ref. \cite{otter2017roadmap}. \par

Rather than focus on the technical aspects of persistent homology, here we give an illustrative example quite similar in spirit to the construction we later present in Sec. \ref{sec:results}. Consider a function $f:[0,1]\to \mathbb{R}$ depicted in Fig. \ref{fig:superlevel}. Our objective is to characterize the structure of $f$; i.e., how the extrema relate to one another. To quantify this structure, we consider the superlevel set $L_{c}^{+}(f)  = \{x \in [0,1]|f(x) \ge c\}$. For any value of $c$ (the dashed lines in Fig. \ref{fig:superlevel}, known as the filtration value) the superlevel set is a set of intervals in $[0,1]$, indicated by the black bars at the bottom of each panel. The number of intervals is the number of connected components (clusters); in topological terms, the Betti number $\beta_{0}$. As we decrease $c$, the number of connected clusters fluctuates until, when $c =c_{min}:= \min_{x}f(x)$, we have $L_{c_{max}}^{+} = [0,1]$, the domain. Along the filtration we can monitor which clusters merge into one another. \par

The end result of capturing the progression of homology groups is a \textit{persistence diagram} showing the filtration value at which a connected component appears in the filtration (known as the \textit{birth} time) on the ordinate, and when a cluster merges with another cluster (the \textit{death time} on the abscissa. Each \textit{persistence point} has a distance from the diagonal: the farther from the diagonal, the longer-lived the cluster was in the filtration. Fig. \ref{fig:superlevel} depicts the persistence diagram with notation indicating the minima and maxima of the functional that identify each cluster. Note that the point $(a,-\infty)$ indicates the largest component, the domain, which technically has infinite lifetime. Next to the persistence diagram in Fig. \ref{fig:superlevel} we depict the dendrogram, or \textit{merge tree} corresponding to the superlevel set filtration: external (internal) nodes denote the maxima (minima). \par 

One remarkable property of persistent homology is its stability with respect to perturbations of the underlying functional. In particular, a series of algebraic stability results imply the Wasserstein metric of two persistence diagrams is continuous with respect to $\ell_{p}$ norms between the underlying filtration functionals \cite{Cohen-Steiner_Edelsbrunner_Harer_2007,Bauer_Lesnick_2015}. Put in simpler terms, small perturbations to the superlevel sets result in small perturbations to the persistence diagrams.\par

\begin{figure}
    \centering
    \includegraphics{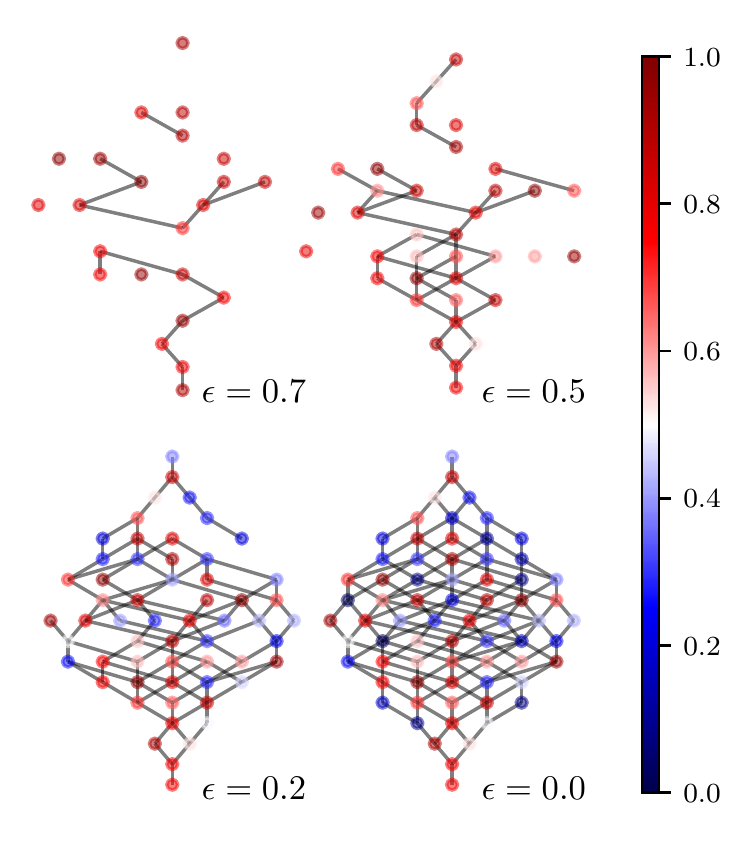}
    \caption{Example of the Fock space $G_{\mathcal{F}}$; colors of nodes denote the value of a random potential. For each value of $\epsilon \in \{0.7,0.5,0.2,0.0\}$ the superlevel set (subgraph) $G[V_{\epsilon}]$ is displayed. }
    \label{fig:network_filtration}
\end{figure}

The unique idea of persistent homology is that the aforementioned pipeline is readily generalizable to higher dimensions. The filtration of intervals generalizes to filtrations of simplicial complexes, and the notion of clusters generalizes to homology groups of a sequence of simplicial complexes. A filtration need not only be superlevel set filtrations of an underlying functional: any filtered sequence of topological spaces (given a total ordering) has a unique representation via a persistence diagram \cite{carlsson2009zigzag}. \par

The toy example described above informs our study of the morphology of the $L_{2}$ landscape. Instead of a functional on a continuous domain, we perform a superlevel set filtration on $G_{\mathcal{F}}$ given the function $u^{(2)}(E_{0})$ on $V(G)$. For each filtration value $c$, the analogue to the superlevel set is the induced subgraph $G[V_{c}]$, where $V_{c}:= \{I \in V(G_{\mathcal{F}})|u_{I}^{(2)}\ge c\}$. Note that the induced subgraph $G[V_{c}]$ includes both the vertex set $V_{c}$ as well as all of the edges $E$ that have both endpoints in $V_{c}$.  \par

While perhaps a less intuitive than a sublevel set filtration, we chose to perform a superlevel filtration for one major reason. The peaks of the $L_{2}$ landscape indicate where eigenstates localize, while the minima demarcate and separate the regions eigenstate near $E_{0}$ localize. If we performed a sublevel filtration, in the extended regime a large fraction of Fock space sites $I$ would be connected into one cluster well before the maxima indicating the localization centers would appear. Upon appearing these maxima would immediately merge into a connected component and therefore not be recorded in the persistence diagram. By performing a superlevel filtration we ensure the localization center feature in the diagrams. \par 

In Fig. \ref{fig:network_filtration} we depict for visual reference a superlevel set filtration on $G_{\mathcal{F}}$ for $N = 8=2N_{e}$ in 1D and with open boundary conditions. Due to the absence of odd-length cycles (the graph is bipartite), our filtration fails to produce any interesting homology in dimensions greater than zero, as any loops generated persistent throughout the rest of the filtration, and no higher order simplices are generated. However, a different interaction term $H_{t}$ will generate higher dimensional homology, as we discuss in Sec. \ref{sec:conclusion}. As noted above, our pipeline for 0D persistent homology corresponds to single-linkage hierarchical clustering (depicted via the merge tree in Fig. \ref{fig:superlevel}).

\par

There are a multitude of other persistent homology construction that we could have attempted in this work.
Of particular interest would be filtrations based on the distance function derived from the Agmon metric \cite{Filoche_Mayboroda_Tao_2021,Balasubramanian_Liao_Galitski_2020}, which has deep theoretical connections to localization bounds. While an Agmon metric for the $L_{2}$ does not currently exist, it is an exciting prospect and the subject of future work.  \par

\section{Results}\label{sec:results}

With the theoretical underpinnings of the many-body localization landscape and the persistent homology pipeline established, we turn our methodology to the prototypical MBL: the disordered Heisenberg Hamiltonian given in Eq. \ref{eq:ham}. Our objective again is to assess whether the $L_{2}$ landscape shows indications of a \textit{morphological} transition upon approaching the MBL phase. To build these indicators we leverage novel topological summaries and observables generated from persistence diagrams and merge trees. Broadly speaking, we find these observables best discriminate the transition when compared across system sizes.\par 

For the Hamiltonian in Eq. \ref{eq:ham} we assume open boundary conditions and performed $1000$ disorder realizations for the $N=[10,14] $ system sizes and over $500$ for the $N=15$ system size. We sampled $W \in [0.1,8.0]$ in intervals of $\delta W =.2$ as well as sampled $W \in \{10,12,14,16,18,20\}$ deep in the MBL phase for comparison. We use exact diagonalization to compute the eigenspectrum for system sizes $N \in [10,15]$. \par

The critical disorder strength for the phase transition in Eq. \ref{eq:ham} is typically given in the range $W_{c}\in [3.5,4]$, though some works place $W_{c}$ at much higher values \cite{Devakul_Singh_2015,morningstar2022avalanches,Doggen_Schindler_Tikhonov_Mirlin_Neupert_Polyakov_Gornyi_2018}. Whether or not the phase transition survives the thermodynamic limit remains an open question, with recent work posited a strong distinction between the MBL \textit{phase} and the finite-size MBL \textit{regime} that can be probed with numerics \cite{morningstar2022avalanches}. Given our limited system sizes we take no position on the stability issue or the absence of mobility edges in the thermodynamic limit in this work, choosing instead to see if morphological indicators coincide with critical values claimed in the literature.\par

Before moving to the persistent homology topological summaries, we briefly digress to discuss the statistics of eigenenergies in Fock space. Given the $L_{2}$ landscape's close correspondence to the local density of states and $G^{\pm}$, the topological summaries explored below probe similar statistics, greatly expounded upon in recent works we now refer to \cite{Logan_Welsh_2019}.\par 

Note that both the on-site energies and eigenenergies $\mathcal{E},E$ share the same $\overline{E}$, as $H_{t} + H_{V}$ is disorder independent. Due to parity and our choice of $V=1$ we have $\overline{E} = -1/4$ for even system sizes. Meanwhile the variance $\mu_{E}^{2}$ goes as $\mathcal{O}(N)$. 

\begin{figure}
    \centering
    \includegraphics{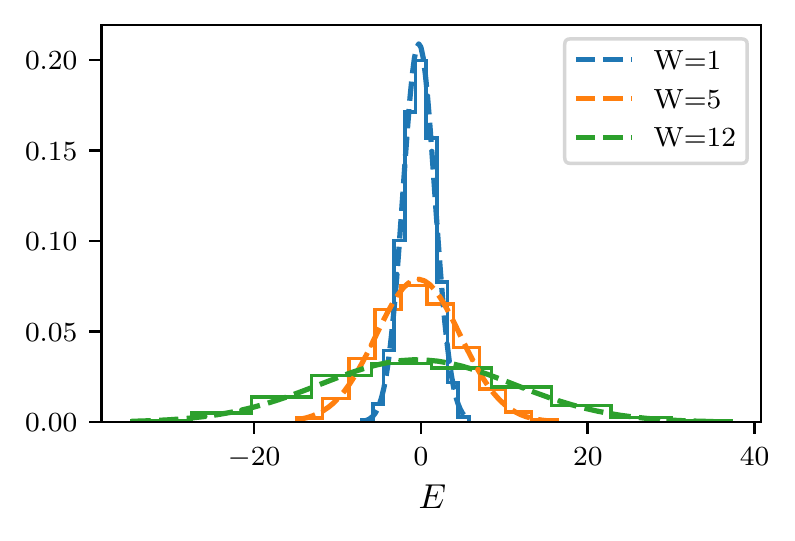}
    \caption{Normalized histogram for $L=15$ of the eigenenergies for $W= (1,5,12)$, the dashed lines show the Gaussian approximation given in Eq. \ref{eq:dos}}
    \label{fig:ex_dos}
\end{figure}

We depict in Fig. \ref{fig:ex_dos} the eigenenergy probability distribution (normalized histogram) for $W \in [1.0,5.0,12.0]$ at $N=15$, as well as the Gaussian approximation to the density of states \begin{align}\label{eq:dos}
    D(\omega) = (2\pi \mu_{E}^{2})^{-1}\exp\left(-(\omega- \overline{E})^{2}/2\mu_{E}^{2}\right),
\end{align} shown as the dashed line for comparison. \par 

\par 

To evaluate $u^{(2)}$, we must select $E_{0}$ and a finite value for $\eta$ which should be smaller than the level spacing near $E_{0}$, but not so small as to impact numerical stability \cite{Herviou_Bardarson_2020}. The choice of $\eta$ dictates the nonlinear extent to which eigenstates far from $E_{0}$ contributes to the $L_{2}$ landscape. To first order Eq. \ref{eq:u2_easy} shifts under $\eta \to \eta + d\eta$ as 
\begin{align}
    (u_{J}^{(2)})^{2} \to (u_{J}^{(2)})^{2} - 2\eta d\eta \sum_{\beta}\frac{|\phi_{J}^{\beta}|^{4}}{|E_{0}+i\eta - E_{\beta}|^{4}}.
\end{align}  
\par 
One natural choice for $\eta$ is the eigenvalue spacing near $E_{0}$, given by  $(N_{\mathcal{H}}D(E_{0}))^{-1}$. Here $D(E_{0})$ is the density of states given in Eq. \ref{eq:dos}, such that $D(E_{0} = \overline{E}) = (\sqrt{2\pi}\mu_{E})$. This then implies $\eta = \sqrt{2\pi}\mu_{E}/N_{\mathcal{H}}$. \par 
However, as noted above, $\mu_{E}$ depends both on $N$ and $W$. Moreover, the level statistics of the model in Eq. \ref{eq:ham} has long been known to shift under the phase transition from GOE (Wegner-Dyson) to Poissonian statistics \cite{Serbyn_Moore_2016}. The precise nature of the level statistics near the transition is still very much controversial \cite{Sierant_Zakrzewski_2019}, and so it is unclear if a more canonical choice for $\eta$ can be made. We make the choice $\eta = \sqrt{2\pi}\mu_{E}/N_{\mathcal{H}}$ for two main reasons.\par  First, the spectral width $\mu_{E}$ is non-parametric in the sense that we do not need to average over some number $k$ level spacing near $\overline{E}$. The number of level spacing necessary for estimation could very well depend upon the disorder strength, rendering this technique possibly biased. \par 
Second, the spectral width $\mu_{E}$ is fundamental to the analysis given in Refs. \cite{Logan_Welsh_2019}, on which we have based our notation and exposition for Fock space spectral statistics. There the authors analyze the stability of the MBL phase in the thermodynamic limit by considering functionals of the resolvent $G_{IJ}^{\pm}$ (particularly, the Feenburg self-energy) under a rescaling $\tilde{E}_{\beta} \to (E_{\beta} - \overline{E})/\mu_{E}$; i.e., the eigenspectrum is normalized to have vanishing mean and a standard deviation of one. Under this rescaling the offset becomes $\tilde{\eta} = \eta/\mu_{E} = \sqrt{2\pi}/N_{\mathcal{H}}$. The rescaling proves central to a mean-field theory approach to the self-consistent determination of the typical self-energy \cite{Logan_Welsh_2019}. Note that $u^{(2)}(\overline{E})$ would be rescaled to $\tilde{u}^{(2)}(0)$ with offset $\tilde{\eta}$, and so we have $\tilde{u}^{(2)}(0) = \mu_{E}u^{(2)}(\overline{E})$. 
Choosing $\eta$ is a nuanced issue and should be the study of future work. \par


\begin{figure}
     \includegraphics[width=\linewidth]{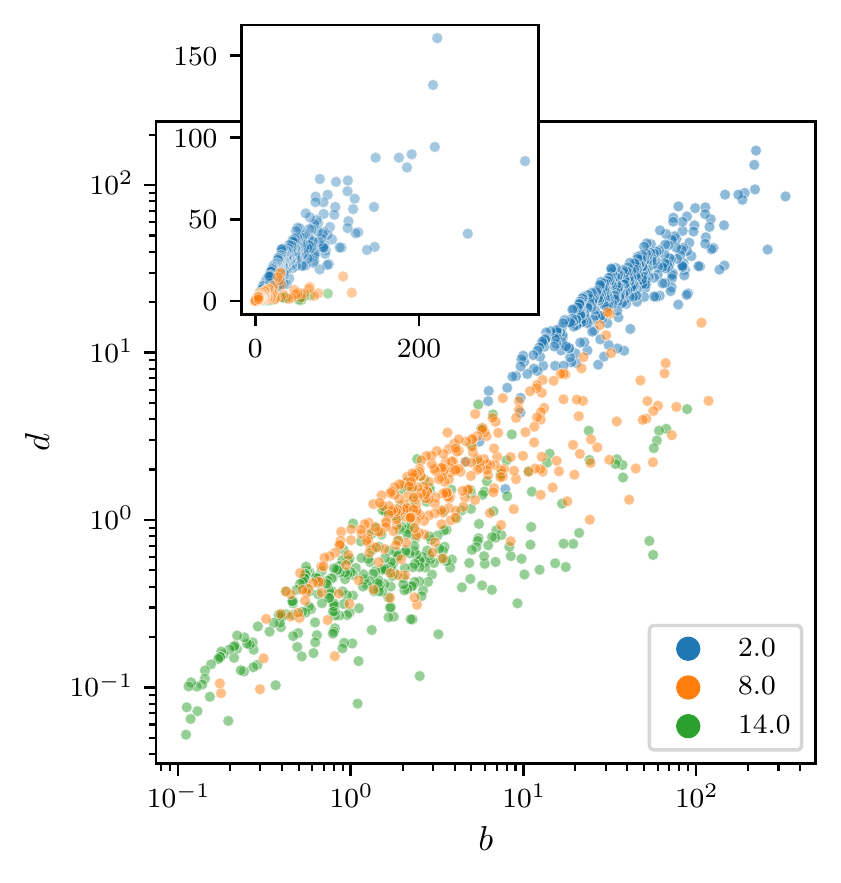}
    \caption{Representative persistence diagrams for disorders strengths $W=\{1.0,3.0,5.0,8.0\}$, plot on a log-log scale. Note that $d < b$ due to the superlevel set filtration. Inset shows the same persistence diagram on a linear scale.}
    \label{fig:ex_pd} 
\end{figure}


\textit{Persistence Diagram -- } Turning to the output of the persistent homology pipeline, Fig. \ref{fig:ex_pd} depicts representative persistence diagrams for the disorder strengths $W = \{2,8,14\}$. Recall that the abscissa and ordinate are the birth and death times, respectively; due to the superlevel set filtration, the death times are smaller than the birth times. We have plotted the persistence diagram in both log-log and linear scale (inset) for comparison. The persistence pairs for the extended regime $W = 2$ cluster at high filtration values with a broad spread of death times. In contrast, in the localized regime the birth and death times are much smaller (corresponding to a small value of $u^{(2)}$). \par 
A perhaps naive interpretation of this behavior is that $|\phi_{J}^{\beta}|^{2} \sim (N_{\mathcal{H}})^{-1}$ on $\mathcal{O}(N_{\mathcal{H}})$ many sites in the extended regime. In contrast, in the localized regime $|\phi_{J}^{\beta}|^{2} \sim N_{\mathcal{H}}^{-\alpha}$ for $\alpha<1$ on $\mathcal{O}(N_{\mathcal{H}}^{\alpha})$ sites. We note that the number of persistence pairs for each diagram corresponds to the total number of maxima and is therefore not constant, a fact that proves consequential when we consider a generalization of the fractal dimension. \par 



\begin{figure}
    \centering
    \includegraphics[width=\linewidth]{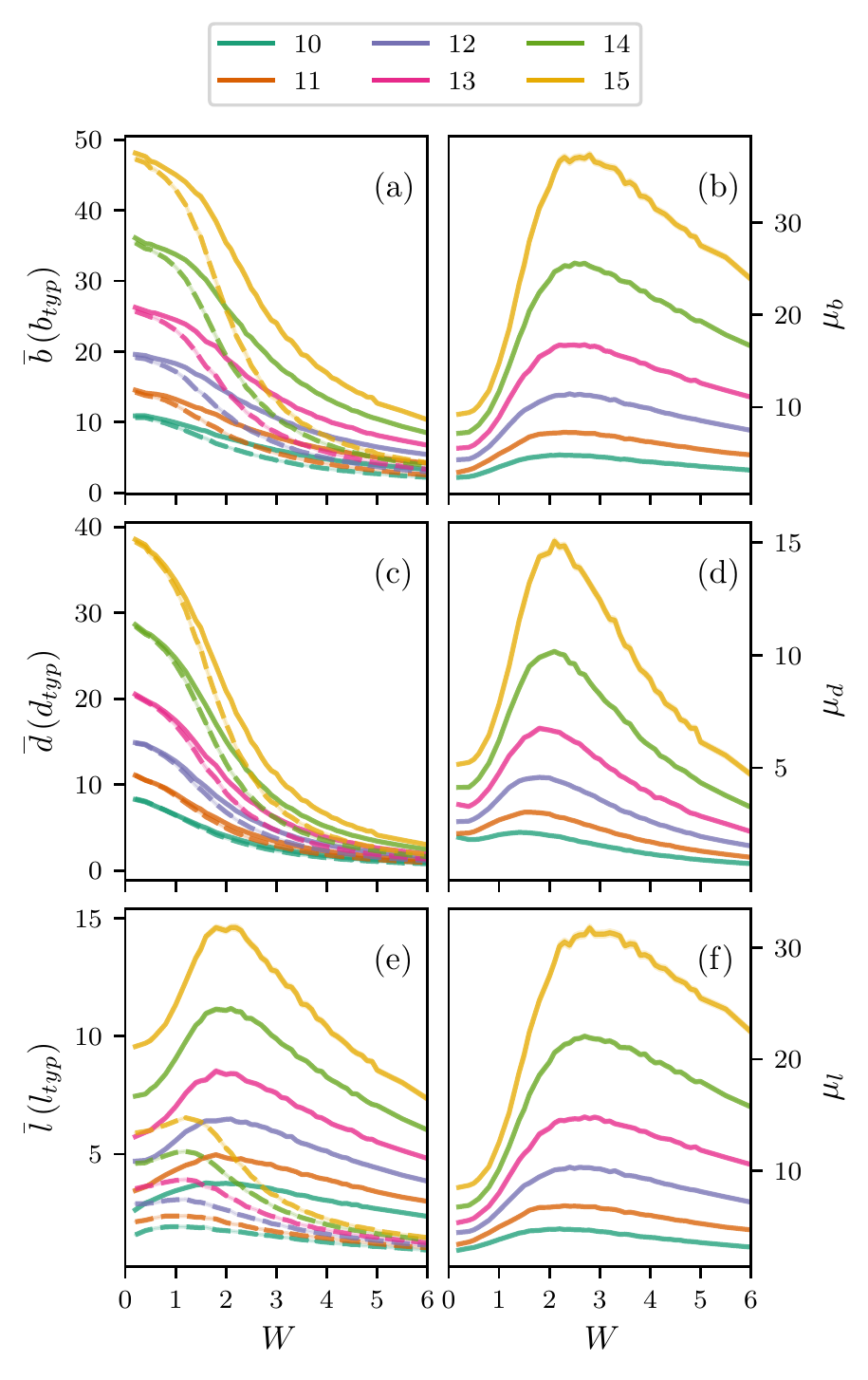}
    \caption{Statistics on the birth, death, and lifetime distributions on the persistence diagrams. (a-b) depict the mean (solid), median (dashed) and standard deviation of the birth values as a function of disorder strength $W$. (c-d) and (e-f) show the corresponding statistics for the death and lifetimes values, respectively.}
    \label{fig:b_d_stats}
\end{figure}


\textit{$(b,d,l)$ statistics -- }
Many topological summaries center on properties of the distributions of the set birth, death, and lifetime values, which we write as vectors $b, \, d, \, l$, respectively. Note that we exclude the persistence pair corresponding to the infinitely long-lived component ($(a,-\infty)$ in Fig. \ref{fig:superlevel}). As we explore in Sec. \ref{sec:lp_norms}, norms of these vectors correspond to observables relevant to percolation theory. \par 
Fig. \ref{fig:b_d_stats}(a) and (b) depicts the arithmetic mean $\overline{b}$, geometric mean $l_{typ}$ (dashed lines), and standard deviation $\mu_{b}$ of the birth distribution against $W$. Fig. \ref{fig:b_d_stats}(c-f) show equivalent data for the death and lifetime distributions. All statistics are calculated within for each disorder realization before taking averaging over realizations. \par 

Plots (a) and (c) show the mean birth and death times decay monotonically with $W$, visually reflecting the change in spread of points in Fig. \ref{fig:ex_pd}. The geometric mean (or typical value) of the birth/death times deviate strongly from the arithmetic mean towards the localized regime, with the deviation more pronounced with larger system sizes and particularly for the birth times. This implies large outliers in the birth times, coincident with the onset of localization centers in the localized regime. In contrast, the difference between arithmetic and geometric mean death times is less pronounced, as the death times (minima of $u^{(2)}$) are lower bounded by zero. \par 
While the arithmetic mean of $l$ is the difference in the mean birth and death times, the same cannot be said for the typical lifetime $l_{typ}$, shown in (e). Here the distinction between arithmetic and geometric mean is even more pronounced, and a peak is observed at $W\approx 1.5$ for $l_{typ}$ and $W \approx 2$ for $\overline{l}$ for the $N=15$ system size. 
\par 
Plot (b), (d), and (f) depict $\mu_{X}$ for $X = (b,d,l)$, respectively, wherein we see peaks generally obey $\mu_{l}^{max} \ge \mu_{b}^{max} > \mu_{d}^{max}$. This follows from the fact that $\mu_{l}$ should add in quadrature, i.e., $\mu_{l}^{2} \approx \mu_{b}^{2} + \mu_{d}^{2}$. The asymmetry in the peak $\mu_{b}, \, \mu_{d}$ values again stems from the fact that both $u^{(2)} \ge 0$ and $d_{i} < b_{i}$ for any persistence pair $(b_{i},d_{i})$. \par

\begin{figure}
    \centering
    \includegraphics[width=\linewidth]{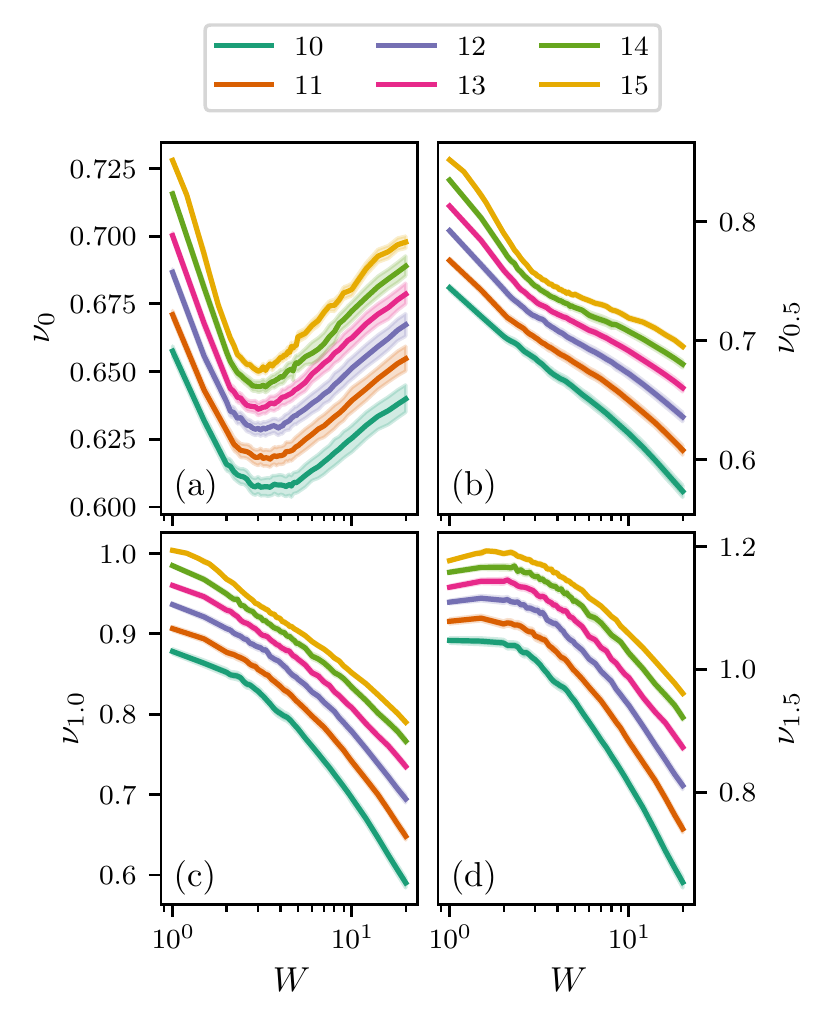}
    \caption{Fractal scaling $\nu_{p}$ for $p = \{0,0.5,1.0,1.5\}$, corresponding to plots (a-d), respectively.}  
    \label{fig:fd_panel}
\end{figure}

\begin{figure}
    \centering
    \includegraphics{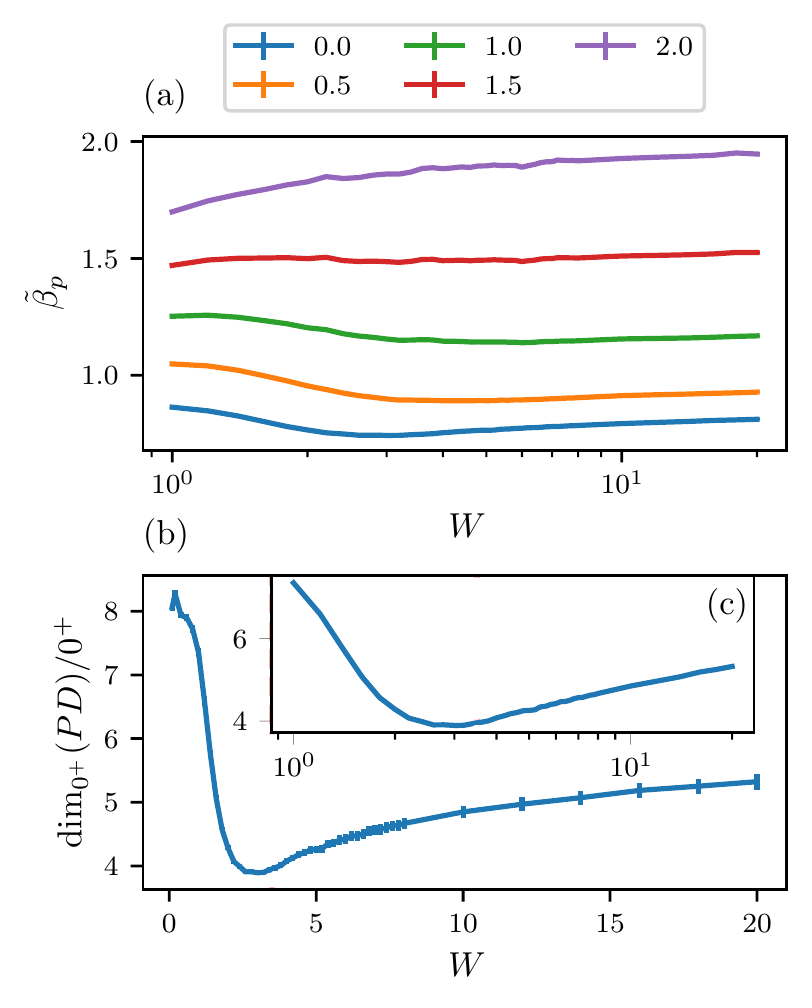}
    \caption{(a) Slope $\tilde{\beta}_{p}$ of $\log ||l||_{p}^{p}$ against $\log N_{\mathcal{H}}$ for $p = (0.0,0.5,1.0,1.5,2.0)$. (b) Fractal dimension $\dim_{p}(PD)$ for the case $p\to 0$. (c) The fractal dimension plotted on a semilog scale.}
    \label{fig:fd_scaling}
\end{figure}


    

\subsubsection{$\ell_{p}$-norms of the persistence diagrams}\label{sec:lp_norms}

We now assess to what extent $\ell_{p}$ norms of the $(b,d,l)$ vectors are sensitive to the phase transition. As we outline below, several of these norms correspond to meaningful quantities in the context of percolation theory, and thus should help shed light on the morphological structure of Fock space under the $L_{2}$ landscape. \par 
\textit{Fractal dimensions -- }
The first observable we consider is how the $\ell_{p}$ norms of $l$ scale with $(N,W)$. We define the function \begin{align}
    \nu_{p}(N,W) = \frac{ \log \langle ||l||_{p}^{p}\rangle}{\log N_{\mathcal{H}}},
\end{align} where $\langle \cdot \rangle $ denotes averaging over disorder realizations holding $W$ fixed. For $p < 1$ $\nu_{p}$ is a semi-norm that emphasizes small-scale homological features. While persistent homology is often used to identify large-scale topological structure, the small-scale structure is relevant in the context of fractality and estimating curvature \cite{Jaquette_Schweinhart_2020,Bubenik_Hull_Patel_Whittle_2020}. Indeed, $\nu_{p}$ was first defined to give a notion of ``persistent fractal dimension" derived from persistent homology and comparable to more traditional measures of fractality like the correlation and box-counting dimensions \cite{Jaquette_Schweinhart_2020}. In its original formulation the persistent fractal dimension is applicable to filtrations built with respect to a finite metric space, such a Vietoris-Rips or C\v ech filtration \cite{Jaquette_Schweinhart_2020}. In our case we apply the same methodology to the $u^{(2)}$ landscape.
We interpret $\nu_{p}$ as quantifying the relative weight of small-scale homological features (short-lived clusters) as a function of $(N,W)$. Note that $||l||_{0}$ counts the number of $u^{(2)}$ maxima. \par We define a persistent fractal dimension as \cite{Jaquette_Schweinhart_2020} \begin{align}
    \text{dim}_{p}(PD) = \frac{p}{1-\tilde{\beta}_{p}}, \\
    \tilde{\beta}_{p} = \lim_{n\to \infty}\sup \nu_{p}.
\end{align} Here $\tilde{\beta}_{p}$ amounts to the slope of $\log  ||l||_{p}^{p}$ against $\log N_{\mathcal{H}}$. Note that the tilde on $\beta$ is to distinguish $\tilde{\beta}_{0}$ from the Betti number $\beta_{0}$, and does not refer to any rescaling such as $u^{(2)} \to \tilde{u}^{(2)}$. \par 
Fig. \ref{fig:fd_panel}(a-d) depicts $\nu_{p}$ for $p \in \{0,0.5,1.0,1.5\}$.
For $p<1$ we observe a noticeable ``kink" in $\nu_{p}$ as $W$ increases, close to $W_{c} \approx 3.5$. For $p>0$ $\nu_{p}$ decreases monotonically with $W$, which upon considering Fig. \ref{fig:b_d_stats}(e) seems contradictory. However, given that $\overline{l} = ||l||_{1}/||l||_{0}$, the peak in Fig. \ref{fig:b_d_stats}(e) stems from the relative decays rates of $\nu_{1}$ versus $\nu_{0}$. \par 

For each $p \in \{0,.5,1.0,1.5,2.0\}$ we calculated $\tilde{\beta}_{p}$, depicted in Fig. \ref{fig:fd_scaling}(a). While $\tilde{\beta}_{p}$ visually appears to scale as $\log W$ in the extended regime, there is a small sub-logarithmic factor. Note that for $p\ge 1/2$ the fractal dimension $\text{dim}_{p}(PD)$ is either divergent at some value of $W$ or negative; however, $\tilde{\beta}_{p} < 0$ for $p=0$. Fig. \ref{fig:fd_scaling}(b) and (c) depict $\text{dim}_{p}(PD)/p$ for $p = 0^{+}$. The inset (c) shows the same plot as (b) but on a log scale.  For $W \ge W_{c}$ $\text{dim}_{0+}(PD)/0^{+} \approx \log W^{a} + b$ with $(a,b) = (.79,3.0)$. \par 

There are several interesting remarks to make concerning the case of $p=0$. First, note that $||l||_{0}$, which we also write as $|PD|$, equates to the total number of persistence pairs in a persistence diagram (hence $|PD|$ denotes cardinality). This quantity is invariant under any monotonically increasing transformation of the underlying filtration; in particular, $|PD|$ is the same whether we use $u^{(2)}$ or $\tilde{u}^{(2)}$. Second, $|PD|$ equates to the number of maxima in the $L_{2}$ landscape. These peaks qualitatively correspond with Fock space sites where eigenstates near $\overline{E}$ also peak. Third, we can compare the scaling behavior of $\nu_{0}$ against the scaling of the Hilbert space dimension $N_{\mathcal{H}}$ against $N$. As noted above, $N_{\mathcal{H}} \propto \exp^{s_{\infty}N}$, where $s_{\infty}$ is the configurational entropy; therefore, $\log N_{\mathcal{H}} \propto s_{\infty}N$. Thus, we could equivalently have computed $\tilde{\beta}_{0}$ by fitting $\log ||l||_{0}$ against $N$, the system size. Then $\tilde{\beta}_{0} \to \tilde{\beta}_{0}/s_{\infty} > 1$ which implies the fractal dimension would become negative. \par 
To try and understand this behavior a bit more analytically, we return to $\nu_{0}$ and consider the behavior of the local density of states $D_{I}(\overline{E}_{0})$. Deep in the ergodic phase the eigenstates can be well-approximated by Gaussian random vectors with vanishing mean and a standard deviation that scales as $N_{\mathcal{H}}^{-1/2}$ \cite{roy2022hilbert}. By time-reversal symmetry the eigenstates are real, and thus the contribution of each eigenstate to the $L_{2}$ landscape is effectively random apart from the denominator, which can also be well-approximated by a Gaussian, see Eq. \ref{eq:dos}. The lack of correlations for $u^{(2)}$ serves as a proxy for a proliferation of local maxima, as neighboring Fock space sites are effectively uncorrelated. As the disorder becomes non-vanishing and the model moves from integrability the eigenstates drift from uncorrelated Gaussians, thus lowering the number of maxima. A more simple picture is that, as the eigenstates begin to localize in regions (clusters of Fock space), the appearance of large amplitude maxima necessitates a decrease in the total number of maxima. \par 
In the other extreme, deep in the MBL phase, we expect eigenstates near $\overline{E}_{0}$ to be tightly clustered around localization centers. Away from these centers, the $L_{2}$ landscape is very small (see Fig. \ref{fig:ex_pd}) and again effectively random, which drives up the number of local maxima. \par 
This qualitative description holds as well for the non-interacting $(V=0)$ case, as deep in the MBL phase the model is perturbatively connected to the non-interacting limit \cite{roy2022hilbert}. 
We can leverage the perturbative connection to the non-interacting limit to partly explain the $\log W$ behavior in $\nu_{p}$. As the non-interacting eigenstates are Slater determinants of the Anderson orbitals, the localization length for an eigenstate scales as $(\log W)^{-1}$ \cite{Scardicchio_Thiery_2017}. Thus the proliferation of uncorrelated regions of the $L_{2}$ landscape should rise in proportion to the decrease in the localization length, which by proxy implies the number of local maxima scale with $\log W$ as observed. Because the number of maxima decrease in the ergodic regime and increase in the MBL regime, there is a natural point between where the number of maxima is minimal, near $W = 3.0$.

\begin{figure}
    \centering
    \includegraphics{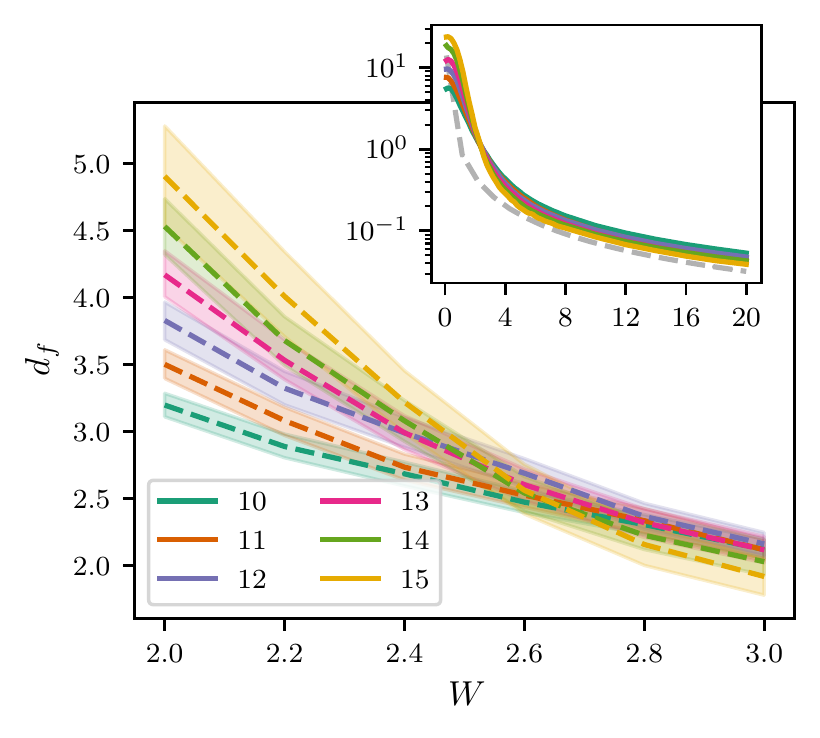}
    \caption{Log of the connectivity threshold $d_{f}$, the last superlevel set wherein at most two connected clusters remain, against $W$. The main plot depict the connectivity threshold near where different system size curves coincide, the inset depicts the curves from $W \in [ 0, 20]$. the dashed black line depicts an the approximate power law scaling $d_{f}\propto W^{-\alpha}$ for $\alpha = 1.15$.}
    \label{fig:conn_thresh}
\end{figure}

\textit{Connectivity threshold -- }
In the context of percolation theory, one figure of merit is the probability of obtaining a giant connected component on Erd\H{o}s-R\'enyi random graphs, wherein a probability is assigned to each edge \cite{bollobas1998random}. For these systems of random graphs a large body of literature focuses on analogues of the MBL localization/delocalization transition with respect to the adjacency matrix \cite{Alt_Ducatez_Knowles_2021,Alt_Knowles_2021}. In the context of our study of the $L_{2}$ landscape morphology, an analogous figure of merit is played by the connectivity threshold $d_{f}$, which we define as the filtration value at which only one connected component remains. By construction, this corresponds to the smallest value of $u^{(2)}$ (global minimum), or equivalently, as $||d||_{-\infty}$. 
Fig. \ref{fig:conn_thresh} depicts the connectivity threshold $d_{f}$ against the $W \in [2.0,3.0]$ wherein we observe the different system sizes cross at various points. The inset shows $d_{f}$ on a log scale across the entire set of $W$ probed. In the localized regime the power law decay goes as $W^{-\alpha}$ with $\alpha \approx 1.15$, which we depict with the dashed black line. The crossing points in the main plot of Fig. \ref{fig:conn_thresh} are again substantially lower than $W_{c}$; it is unclear if the reduced finite size effects at larger values of $N$ would resolve these crossing points. \par


\textit{Maximum spanning tree -- } Also related to percolation theory is the notion of maximum (MaxST) and minimum spanning trees (MinST). On a graph $G= (V,E)$ with edge weight function $w: E \to \mathbb{R} $, the MaxST (MinST) (if it exists, else a spanning forest) is an acyclic connected subgraph such that the sum of the edge weights is maximal (minimal). The weight of the MaxST corresponds to the sum of the death times $||d||_{1}$ of the persistent diagram and is therefore equivalent to the integral of the reduced Betti number with respect to the filtration \cite{Skraba_Thoppe_Yogeshwaran_2020}. The reduced Betti number at filtration value $c$ is specifically the rank of the reduced homology group $H_{0}(G[V_{c}],I_{max})$, where $I_{max}: = \text{argmax}_{V(G)} u_{I}^{(2)}$. Put more plainly, the reduced Betti number is generally one less than the regular Betti number.
\par 

\begin{figure}
    \centering
    \includegraphics[width=\linewidth]{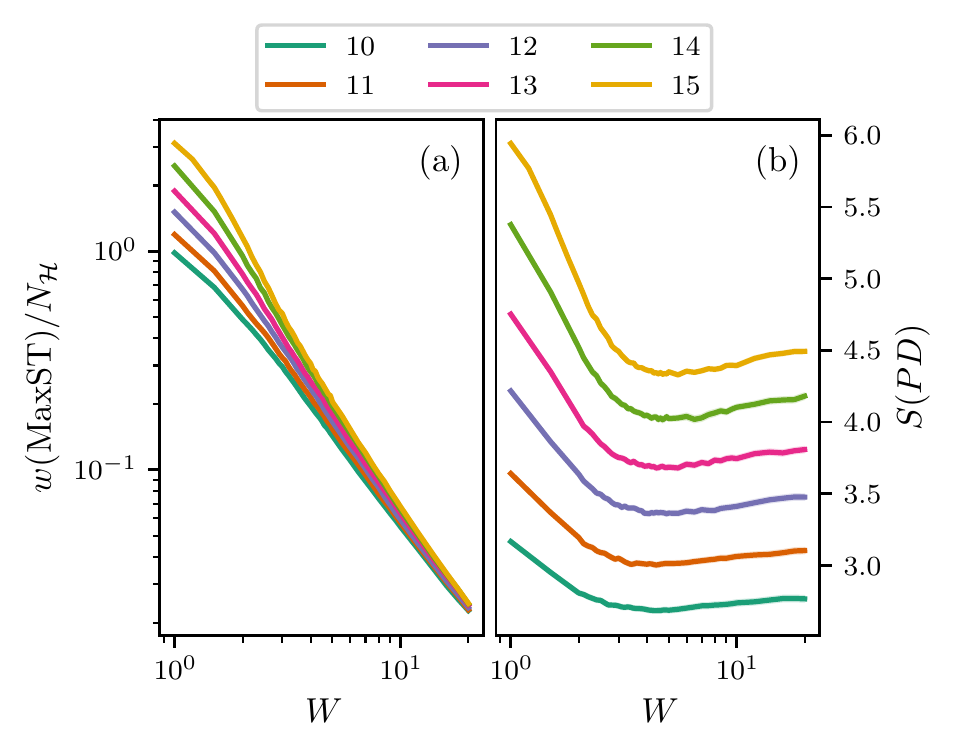}
    \caption{(a) the weight of the maximal spanning tree $w(\text{MaxST})$ against $W$, normalized by $N_{\mathcal{H}}$. (b) The persistent entropy $S(PD)$ against $W$.}
    \label{fig:mean_mst}
\end{figure}

Fig. \ref{fig:mean_mst} (a) depicts  $ w(\text{MaxST})/N_{\mathcal{H}}$ against $W$ on a log-log scale, whereby the power law scaling in the localized regime is manifest. The number of terms in $w(\text{MaxST})$ scales with $|PD| \propto \log W$, which implies the average death time scales as $W^{\gamma}/\log W$ for some scaling exponent $\gamma$ that varies slightly for different system sizes.

\begin{figure}
    \centering
    \includegraphics{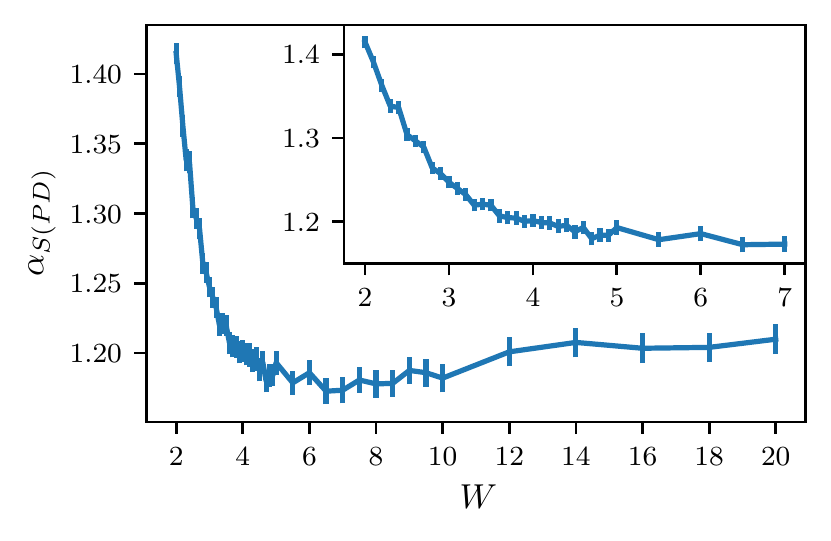}
    \caption{The exponent $\alpha_{S(PD)}$ from fitting $S(PD)\propto N^{\alpha}$. Inset shows the range $W \in [2,7]$.}
    \label{fig:pers_ent_scaling}
\end{figure}

\textit{Persistent entropy --} We turn now to an observable that closely mirrors the participation entropy in accounting both for cardinality and magnitude of persistence pairs: the persistent entropy.
By normalizing $l$ to a probability vector $\tilde{l}:= l/||l||_{1}$, we define the persistent entropy as \begin{align}
    S(PD): = -S(\tilde{l}),
\end{align} where $S(\cdot)$ as before is the entropy function \cite{myers2019persistent}. We can equate $S(PD)$ to a functional of the $\ell_{p}$ norms: \begin{align}
    S(PD) = \frac{S(l)}{||l||_{1}} - \sum_{i}\ln l_{i}.
\end{align}

The persistent entropy has been extensively used in the context of detecting both classical and quantum phase transitions \cite{Tran_Chen_Hasegawa_2021}. Qualitatively, for a fixed value $|PD|$ a large $S(PD)$ implies the lifetimes are largely constant. In contrast, a small $S(PD)$ implies the prevalence of vanishing few clusters with lifetime disproportionately large. Clearly $S(PD) \le \log |PD|$, just as the Schmidt rank bounds the entanglement entropy and $\log N_{\mathcal{H}}$ bounds the participation entropies.\par

Fig. \ref{fig:mean_mst}(b) depicts the persistent entropy $S(PD)$ as a function of $W$. In contrast to the observables examined above, $S(PD)$ flattens out to nearly constant, with a very slow, sublogarithmic scaling in $ W$. Qualitatively, near constant $S(PD)$ is analogous to the area law entanglement entropy exhibited by localized states. Heuristically, the lifetimes of the Fock site clusters indicating eigenstates near $\overline{E}$ roughly correspond to the large Schmidt coefficients of a localized eigenstate under a bipartition. We expect a relatively large contribution to the $L^{2}$ landscape from these localization centers while the rest of the landscape is vanishingly small. There is a competition between the number of local maxima (increasing as $\log W$) versus the relative persistence of these maxima. Fig. \ref{fig:mean_mst} indicates that the maxima proliferation outweighs the disproportional lifetime of these maxima in slowly driving up the persistent entropy. \par 

\par 
Just as $S(\rho_{A})$ scales with $N$ in the extended regime, so too does $S(PD)$; in Fig. \ref{fig:pers_ent_scaling} we depict the coefficient $\alpha$ from a fit $S(PD) \propto N^{\alpha}$, with inset zoomed into the region $W\in [2.0,7.0]$. The coefficient $\alpha$ plateaus in the region $W\in [4,6]$ before slowly increasing with $W$. This behavior is very similar to Fig. \ref{fig:fd_scaling} (b) which depicted the fractal dimension, though the minimum value for $\alpha$ seems be at a higher $W$ and the scaling is not quite logarithmic.

\begin{figure}
    \centering  
    \includegraphics[]{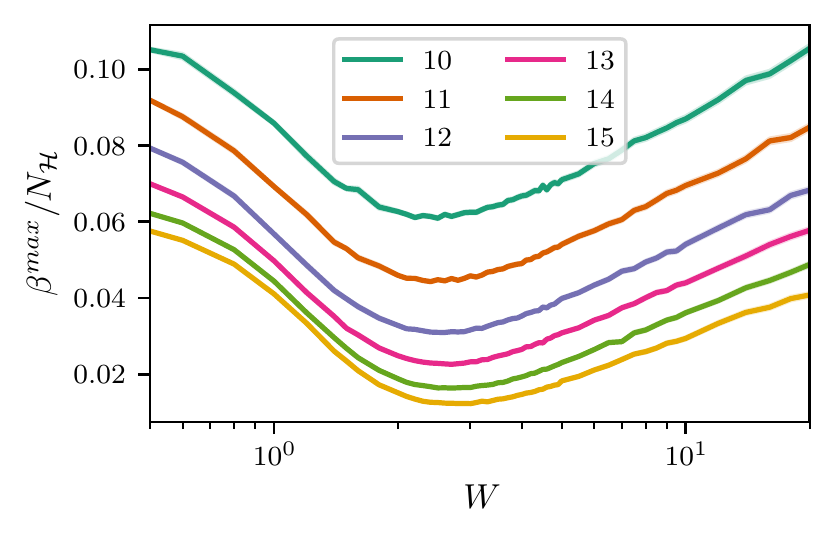}
    \caption{The peak value of $\beta_{0}$ during the filtration (equivalently, the maximum number of independent clusters at any point during the filtration), normalized by $N_{\mathcal{H}}$. Note that $\beta_{0}^{max}$ goes as $\log W$ in the localized regime.}
    \label{fig:betti_max}
\end{figure}

\begin{figure}
    \centering
    \includegraphics{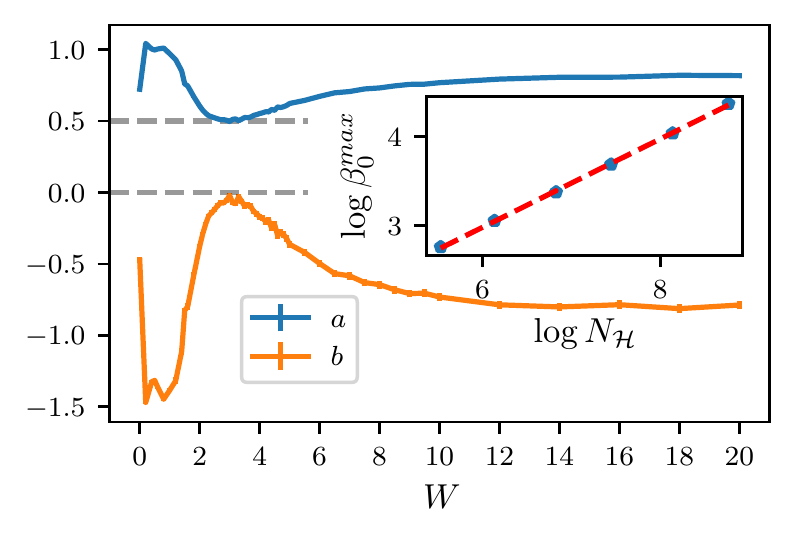}
    \caption{ Scaling exponents from fitting $\beta_{0}^{max} \propto N_{\mathcal{H}}^{a}\left(\log N_{\mathcal{H}}\right)^{b}$. 
    The horizontal dashed lines at $0$ and $.5$ roughly correspond to the maximum (minimum) value of $b$ ($a$), which occurs at $W \approx 3.0$. Inset depicts representative fitting of $\beta_{0}^{max}$ against $N_{\mathcal{H}}$ at $W = 3.0$.
    }
    \label{fig:betti_max_fitting}
\end{figure}

\textit{Maximum Betti number --}
The final observable we consider is the maximum value of the Betti number $\beta_{0}^{max}$ encountered during the superlevel set filtration. This observable has been leveraged in recent works as an indicator for quantum phase transitions \cite{Olsthoorn_Balatsky_2021}. Fig. \ref{fig:betti_max} depicts $\beta_{0}^{max}/N_{\mathcal{H}}$ against $W$ wherein we see that, quite similar to $|PD|$, a logarithmic growth in the extended regime. Appealing to the heuristical argument given above, in the extended regime $\beta_{0}^{max}$ occurs very late in the filtration, growing in proportional to the extent of decorrelated Fock space. Fig. \ref{fig:betti_max_fitting} depicts the scaling of $\beta_{0}^{max} \propto N_{\mathcal{H}}^{a}\left(\log N_{\mathcal{H}}\right)^{b}$ with exponents $(a,b)$ disorder-dependent. Intriguingly, at $W \approx 3.0$ $(a,b) \approx (.5,0)$, which implies $\beta_{0}^{max} \propto \sqrt{N_{\mathcal{H}}}$. This is roughly the same disorder value for which the minimum in $\text{dim}_{0+}(PD)$ occurred. Though it is at present unclear if the extremal behavior of the scaling exponents $(a,b)$ would shift for larger values of $N$, the simplicity of the scaling at $W \approx 3.0$ indicates deeper behavior that merits exploration in future work.

\section{Conclusion}\label{sec:conclusion}
This work takes an exploratory approach to leveraging persistent homology as a descriptor for the complex correlational structure of eigenstates in Fock space. We utilize a new construction, the $L_{2}$ landscape, to procure topological summaries of the mid-spectrum Fock space structure. \par 

Overall, we found that several observables, including the cardinality $|PD|$ of the persistence diagrams (Fig. \ref{fig:fd_scaling}), the persistent entropy (Fig. \ref{fig:pers_ent_scaling}) and the peak Betti number (Fig. \ref{fig:betti_max})  all exhibit scaling in either $N$ or $N_{\mathcal{H}}$ near $W_{c}$ indicative of a transition. This is especially evident for $|PD|$ which is perhaps the most easily interpreted observable presented here, as it corresponds to the number of maxima of an approximate local density of states. \par 
Along with these observables we proposed a novel notion of a fractal dimension readily applicable to other functionals defined on Fock space, including but not limited to self-energies and the local density of states. While a recent study expanded the participation entropy definition to a normalized version of the local density of states \cite{Murphy_Wortis_Atkinson_2011}, our approach is fundamentally different. The notion of fractal dimension presented here incorporates not only the Fock space lattice but also the complex morphology of level sets. \par 
Still other observables painted a less intuitive picture of the MBL phase transition, and it should be the aim of future work to discriminate observables that inaccurately pinpoint the phase transition due to finite-size effects from observables that do not probe the transition at all. 
Two possible indicators for the transition, the connectivity threshold $d_{f}$ (Fig. \ref{fig:conn_thresh}) and the $(b,d,l)$ statistics (Fig. \ref{fig:b_d_stats}), show substantial drift in the crossing of different system size curves, indicative of the finite-size effects at the relatively small system sizes probed here. \par 
Throughout this work we have posed a heuristical understanding that is largely based upon the local density of states which, by Eq. \ref{eq:urho_conn}, the $L_{2}$ is connected to. However, lacking a deeper, more analytic understanding of how $u^{(2)}$ evolves with $W$ limits the scope to which a topological data analysis lens can shed light on the transition. Much of this limitation stems from the difficulty in interpreting persistent homology in terms of standard statistical observables. While persistent homology is strongly connected to discrete Morse theory \cite{saucan2020discrete}, to our knowledge there exists a limited amount of work in the topological data analysis literature to bridge homological and statistical observables. We conclude with several proposed future directions to take this exploratory analysis, as the use of persistent homology in the context of quantum information and many-body systems is still nascent. \par 
\textit{Persistent Homology for functionals -- } The PH pipeline explored here can equally be applied to individual eigenstate as performed in Ref. \cite{He_Xia_Angelakis_Song_Chen_Leykam_2022} for the single-body Aubry-Andr\'e model.
In fact, the persistent homology pipeline could be readily applied to any functional represented on $G_{\mathcal{F}}$, including the local density of states and the self-energies. The edges $E(G_{\mathcal{F}})$ can have filtration values that differ from their boundary nodes. \par
As noted in Sec. \ref{sec:theory}, the original $LL$ gives rise to a degenerate Agmon metric on $G_{\mathcal{F}}$. From this metric a series of localization theorems limit the extent eigenstate support extended into ``classically disallowed regions" \cite{Filoche_Mayboroda_Tao_2021}. This Agmon metric determines a filtration on the complete graph $G$ with vertex set $V(G)$ the Fock space sites. However, a complete graph would become computationally prohibitive with large $N$, at which point approximate methods such as witness complexes could be used \cite{de2004topological}.
 
\textit{Higher-dimensional Homology -- }
The analysis given in this work is restricted to 0D homology due to the interaction graph $G_{\mathcal{F}}$: the bipartiteness of the graph precludes any $2$-simplices from forming. This obstruction can, of course, be overcome by simply choosing a different Hamiltonian; e.g., including a $\sigma_{x}$ term that breaks total magnetization conservation. Under this inclusion persistent homology could be used to study the occurrence of higher-dimensional homology features like cycles and voids.\par 

\textit{Approximations for larger system sizes -- }
For larger system sizes exact diagonalization is computationally expensive. Thus, approximate forms of the $L_{2}$ landscape could be computed via recursive approximations to the diagonal of the Green function \cite{Herviou_Bardarson_2020}. The original LL equation in Eq. \ref{eq:ll} can be written as a Rayleigh quotient and is therefore amenable to DMRG and tensor network approaches. Recent tensor network representations of Green functions could be leveraged to scale the $u^{(2)}$ calculation \cite{shinaoka2020sparse}. \par

\begin{acknowledgments}
GH acknowledges useful conversations with Felix Leditzky, Di Luo, Yuliy Baryshnikov, Shmuel Weinberger, Giuseppe De Tomasi, and Henry Adams. This work made use of the Illinois Campus Cluster, a computing resource that is operated by the Illinois Campus Cluster Program (ICCP) in conjunction with the National Center for Supercomputing Applications (NCSA) and which is supported by funds from the University of Illinois at Urbana-Champaign. We acknowledge support from the Department of Energy grant DOE DESC0020165.
\end{acknowledgments}
\bibliography{library_fst.bib}

\begin{thebibliography}{73}%
\makeatletter
\providecommand \@ifxundefined [1]{%
 \@ifx{#1\undefined}
}%
\providecommand \@ifnum [1]{%
 \ifnum #1\expandafter \@firstoftwo
 \else \expandafter \@secondoftwo
 \fi
}%
\providecommand \@ifx [1]{%
 \ifx #1\expandafter \@firstoftwo
 \else \expandafter \@secondoftwo
 \fi
}%
\providecommand \natexlab [1]{#1}%
\providecommand \enquote  [1]{``#1''}%
\providecommand \bibnamefont  [1]{#1}%
\providecommand \bibfnamefont [1]{#1}%
\providecommand \citenamefont [1]{#1}%
\providecommand \href@noop [0]{\@secondoftwo}%
\providecommand \href [0]{\begingroup \@sanitize@url \@href}%
\providecommand \@href[1]{\@@startlink{#1}\@@href}%
\providecommand \@@href[1]{\endgroup#1\@@endlink}%
\providecommand \@sanitize@url [0]{\catcode `\\12\catcode `\$12\catcode
  `\&12\catcode `\#12\catcode `\^12\catcode `\_12\catcode `\%12\relax}%
\providecommand \@@startlink[1]{}%
\providecommand \@@endlink[0]{}%
\providecommand \url  [0]{\begingroup\@sanitize@url \@url }%
\providecommand \@url [1]{\endgroup\@href {#1}{\urlprefix }}%
\providecommand \urlprefix  [0]{URL }%
\providecommand \Eprint [0]{\href }%
\providecommand \doibase [0]{http://dx.doi.org/}%
\providecommand \selectlanguage [0]{\@gobble}%
\providecommand \bibinfo  [0]{\@secondoftwo}%
\providecommand \bibfield  [0]{\@secondoftwo}%
\providecommand \translation [1]{[#1]}%
\providecommand \BibitemOpen [0]{}%
\providecommand \bibitemStop [0]{}%
\providecommand \bibitemNoStop [0]{.\EOS\space}%
\providecommand \EOS [0]{\spacefactor3000\relax}%
\providecommand \BibitemShut  [1]{\csname bibitem#1\endcsname}%
\let\auto@bib@innerbib\@empty
\bibitem [{\citenamefont {Abanin}\ \emph {et~al.}(2019)\citenamefont {Abanin},
  \citenamefont {Altman}, \citenamefont {Bloch},\ and\ \citenamefont
  {Serbyn}}]{Abanin_Altman_Bloch_Serbyn_2019}%
  \BibitemOpen
  \bibfield  {author} {\bibinfo {author} {\bibfnamefont {D.~A.}\ \bibnamefont
  {Abanin}}, \bibinfo {author} {\bibfnamefont {E.}~\bibnamefont {Altman}},
  \bibinfo {author} {\bibfnamefont {I.}~\bibnamefont {Bloch}}, \ and\ \bibinfo
  {author} {\bibfnamefont {M.}~\bibnamefont {Serbyn}},\ }\href {\doibase
  10.1103/RevModPhys.91.021001} {\bibfield  {journal} {\bibinfo  {journal}
  {Reviews of Modern Physics}\ }\textbf {\bibinfo {volume} {91}},\ \bibinfo
  {pages} {021001} (\bibinfo {year} {2019})}\BibitemShut {NoStop}%
\bibitem [{\citenamefont {Thomson}\ and\ \citenamefont
  {Schiro}()}]{Thomson_Schiro}%
  \BibitemOpen
  \bibfield  {author} {\bibinfo {author} {\bibfnamefont {S.~J.}\ \bibnamefont
  {Thomson}}\ and\ \bibinfo {author} {\bibfnamefont {M.}~\bibnamefont
  {Schiro}},\ }\href@noop {} {\bibinfo  {journal} {SciPost Physics}\ ,\
  \bibinfo {pages} {39}}\BibitemShut {NoStop}%
\bibitem [{\citenamefont {Scardicchio}\ and\ \citenamefont
  {Thiery}(2017)}]{Scardicchio_Thiery_2017}%
  \BibitemOpen
\bibfield  {journal} {  }\bibfield  {author} {\bibinfo {author} {\bibfnamefont
  {A.}~\bibnamefont {Scardicchio}}\ and\ \bibinfo {author} {\bibfnamefont
  {T.}~\bibnamefont {Thiery}},\ }\href {http://arxiv.org/abs/1710.01234}
  {\bibfield  {journal} {\bibinfo  {journal} {arXiv:1710.01234 [cond-mat]}\ }
  (\bibinfo {year} {2017})},\ \bibinfo {note} {arXiv: 1710.01234}\BibitemShut
  {NoStop}%
\bibitem [{\citenamefont {Alet}\ and\ \citenamefont
  {Laflorencie}(2018{\natexlab{a}})}]{alet2018many}%
  \BibitemOpen
  \bibfield  {author} {\bibinfo {author} {\bibfnamefont {F.}~\bibnamefont
  {Alet}}\ and\ \bibinfo {author} {\bibfnamefont {N.}~\bibnamefont
  {Laflorencie}},\ }\href@noop {} {\bibfield  {journal} {\bibinfo  {journal}
  {Comptes Rendus Physique}\ }\textbf {\bibinfo {volume} {19}},\ \bibinfo
  {pages} {498} (\bibinfo {year} {2018}{\natexlab{a}})}\BibitemShut {NoStop}%
\bibitem [{\citenamefont {Gopalakrishnan}\ \emph {et~al.}(2015)\citenamefont
  {Gopalakrishnan}, \citenamefont {M{\"u}ller}, \citenamefont {Khemani},
  \citenamefont {Knap}, \citenamefont {Demler},\ and\ \citenamefont
  {Huse}}]{gopalakrishnan2015low}%
  \BibitemOpen
  \bibfield  {author} {\bibinfo {author} {\bibfnamefont {S.}~\bibnamefont
  {Gopalakrishnan}}, \bibinfo {author} {\bibfnamefont {M.}~\bibnamefont
  {M{\"u}ller}}, \bibinfo {author} {\bibfnamefont {V.}~\bibnamefont {Khemani}},
  \bibinfo {author} {\bibfnamefont {M.}~\bibnamefont {Knap}}, \bibinfo {author}
  {\bibfnamefont {E.}~\bibnamefont {Demler}}, \ and\ \bibinfo {author}
  {\bibfnamefont {D.~A.}\ \bibnamefont {Huse}},\ }\href@noop {} {\bibfield
  {journal} {\bibinfo  {journal} {Physical Review B}\ }\textbf {\bibinfo
  {volume} {92}},\ \bibinfo {pages} {104202} (\bibinfo {year}
  {2015})}\BibitemShut {NoStop}%
\bibitem [{\citenamefont {Kim}\ \emph {et~al.}(2014)\citenamefont {Kim},
  \citenamefont {Chandran},\ and\ \citenamefont {Abanin}}]{kim2014local}%
  \BibitemOpen
  \bibfield  {author} {\bibinfo {author} {\bibfnamefont {I.~H.}\ \bibnamefont
  {Kim}}, \bibinfo {author} {\bibfnamefont {A.}~\bibnamefont {Chandran}}, \
  and\ \bibinfo {author} {\bibfnamefont {D.~A.}\ \bibnamefont {Abanin}},\
  }\href@noop {} {\bibfield  {journal} {\bibinfo  {journal} {arXiv preprint
  arXiv:1412.3073}\ } (\bibinfo {year} {2014})}\BibitemShut {NoStop}%
\bibitem [{\citenamefont {Pekker}\ and\ \citenamefont
  {Clark}(2017)}]{Pekker_Clark_2017}%
  \BibitemOpen
  \bibfield  {author} {\bibinfo {author} {\bibfnamefont {D.}~\bibnamefont
  {Pekker}}\ and\ \bibinfo {author} {\bibfnamefont {B.~K.}\ \bibnamefont
  {Clark}},\ }\href {\doibase 10.1103/PhysRevB.95.035116} {\bibfield  {journal}
  {\bibinfo  {journal} {Physical Review B}\ }\textbf {\bibinfo {volume} {95}},\
  \bibinfo {pages} {035116} (\bibinfo {year} {2017})}\BibitemShut {NoStop}%
\bibitem [{\citenamefont {Alet}\ and\ \citenamefont
  {Laflorencie}(2018{\natexlab{b}})}]{Alet_Laflorencie_2018}%
  \BibitemOpen
  \bibfield  {author} {\bibinfo {author} {\bibfnamefont {F.}~\bibnamefont
  {Alet}}\ and\ \bibinfo {author} {\bibfnamefont {N.}~\bibnamefont
  {Laflorencie}},\ }\href {\doibase 10.1016/j.crhy.2018.03.003} {\bibfield
  {journal} {\bibinfo  {journal} {Comptes Rendus Physique}\ }\textbf {\bibinfo
  {volume} {19}},\ \bibinfo {pages} {498–525} (\bibinfo {year}
  {2018}{\natexlab{b}})},\ \bibinfo {note} {arXiv:1711.03145
  [cond-mat]}\BibitemShut {NoStop}%
\bibitem [{\citenamefont {De~Roeck}\ and\ \citenamefont
  {Imbrie}(2017)}]{Roeck_Imbrie_2017}%
  \BibitemOpen
  \bibfield  {author} {\bibinfo {author} {\bibfnamefont {W.}~\bibnamefont
  {De~Roeck}}\ and\ \bibinfo {author} {\bibfnamefont {J.~Z.}\ \bibnamefont
  {Imbrie}},\ }\href {\doibase 10.1098/rsta.2016.0422} {\bibfield  {journal}
  {\bibinfo  {journal} {Philosophical Transactions of the Royal Society A:
  Mathematical, Physical and Engineering Sciences}\ }\textbf {\bibinfo {volume}
  {375}},\ \bibinfo {pages} {20160422} (\bibinfo {year} {2017})}\BibitemShut
  {NoStop}%
\bibitem [{\citenamefont {Imbrie}\ \emph {et~al.}(2017)\citenamefont {Imbrie},
  \citenamefont {Ros},\ and\ \citenamefont
  {Scardicchio}}]{Imbrie_Ros_Scardicchio_2017}%
  \BibitemOpen
  \bibfield  {author} {\bibinfo {author} {\bibfnamefont {J.~Z.}\ \bibnamefont
  {Imbrie}}, \bibinfo {author} {\bibfnamefont {V.}~\bibnamefont {Ros}}, \ and\
  \bibinfo {author} {\bibfnamefont {A.}~\bibnamefont {Scardicchio}},\ }\href
  {\doibase 10.1002/andp.201600278} {\bibfield  {journal} {\bibinfo  {journal}
  {Annalen der Physik}\ }\textbf {\bibinfo {volume} {529}},\ \bibinfo {pages}
  {1600278} (\bibinfo {year} {2017})},\ \bibinfo {note} {arXiv:
  1609.08076}\BibitemShut {NoStop}%
\bibitem [{\citenamefont {Luitz}\ \emph {et~al.}(2015)\citenamefont {Luitz},
  \citenamefont {Laflorencie},\ and\ \citenamefont
  {Alet}}]{Luitz_Laflorencie_Alet_2015}%
  \BibitemOpen
  \bibfield  {author} {\bibinfo {author} {\bibfnamefont {D.~J.}\ \bibnamefont
  {Luitz}}, \bibinfo {author} {\bibfnamefont {N.}~\bibnamefont {Laflorencie}},
  \ and\ \bibinfo {author} {\bibfnamefont {F.}~\bibnamefont {Alet}},\ }\href
  {\doibase 10.1103/PhysRevB.91.081103} {\bibfield  {journal} {\bibinfo
  {journal} {Physical Review B}\ }\textbf {\bibinfo {volume} {91}},\ \bibinfo
  {pages} {081103} (\bibinfo {year} {2015})},\ \bibinfo {note} {arXiv:
  1411.0660}\BibitemShut {NoStop}%
\bibitem [{\citenamefont {Cookmeyer}\ \emph {et~al.}(2020)\citenamefont
  {Cookmeyer}, \citenamefont {Motruk},\ and\ \citenamefont
  {Moore}}]{Cookmeyer_Motruk_Moore_2020}%
  \BibitemOpen
  \bibfield  {author} {\bibinfo {author} {\bibfnamefont {T.}~\bibnamefont
  {Cookmeyer}}, \bibinfo {author} {\bibfnamefont {J.}~\bibnamefont {Motruk}}, \
  and\ \bibinfo {author} {\bibfnamefont {J.~E.}\ \bibnamefont {Moore}},\ }\href
  {\doibase 10.1103/PhysRevB.101.174203} {\bibfield  {journal} {\bibinfo
  {journal} {Physical Review B}\ }\textbf {\bibinfo {volume} {101}},\ \bibinfo
  {pages} {174203} (\bibinfo {year} {2020})}\BibitemShut {NoStop}%
\bibitem [{\citenamefont {Schulz}\ \emph {et~al.}(2019)\citenamefont {Schulz},
  \citenamefont {Hooley}, \citenamefont {Moessner},\ and\ \citenamefont
  {Pollmann}}]{Schulz_Hooley_Moessner_Pollmann_2019}%
  \BibitemOpen
  \bibfield  {author} {\bibinfo {author} {\bibfnamefont {M.}~\bibnamefont
  {Schulz}}, \bibinfo {author} {\bibfnamefont {C.~A.}\ \bibnamefont {Hooley}},
  \bibinfo {author} {\bibfnamefont {R.}~\bibnamefont {Moessner}}, \ and\
  \bibinfo {author} {\bibfnamefont {F.}~\bibnamefont {Pollmann}},\ }\href
  {\doibase 10.1103/PhysRevLett.122.040606} {\bibfield  {journal} {\bibinfo
  {journal} {Physical Review Letters}\ }\textbf {\bibinfo {volume} {122}},\
  \bibinfo {pages} {040606} (\bibinfo {year} {2019})},\ \bibinfo {note} {arXiv:
  1808.01250}\BibitemShut {NoStop}%
\bibitem [{\citenamefont {Morningstar}\ \emph {et~al.}(2020)\citenamefont
  {Morningstar}, \citenamefont {Huse},\ and\ \citenamefont
  {Imbrie}}]{Morningstar_Huse_Imbrie_2020}%
  \BibitemOpen
  \bibfield  {author} {\bibinfo {author} {\bibfnamefont {A.}~\bibnamefont
  {Morningstar}}, \bibinfo {author} {\bibfnamefont {D.~A.}\ \bibnamefont
  {Huse}}, \ and\ \bibinfo {author} {\bibfnamefont {J.~Z.}\ \bibnamefont
  {Imbrie}},\ }\href {\doibase 10.1103/PhysRevB.102.125134} {\bibfield
  {journal} {\bibinfo  {journal} {Physical Review B}\ }\textbf {\bibinfo
  {volume} {102}},\ \bibinfo {pages} {125134} (\bibinfo {year}
  {2020})}\BibitemShut {NoStop}%
\bibitem [{\citenamefont {Monteiro}\ \emph {et~al.}()\citenamefont {Monteiro},
  \citenamefont {Micklitz}, \citenamefont {Tezuka},\ and\ \citenamefont
  {Altland}}]{Monteiro_Micklitz_Tezuka_Altland}%
  \BibitemOpen
  \bibfield  {author} {\bibinfo {author} {\bibfnamefont {F.}~\bibnamefont
  {Monteiro}}, \bibinfo {author} {\bibfnamefont {T.}~\bibnamefont {Micklitz}},
  \bibinfo {author} {\bibfnamefont {M.}~\bibnamefont {Tezuka}}, \ and\ \bibinfo
  {author} {\bibfnamefont {A.}~\bibnamefont {Altland}},\ }\href@noop {} {\ ,\
  \bibinfo {pages} {21}}\BibitemShut {NoStop}%
\bibitem [{\citenamefont {Vidmar}\ \emph {et~al.}(2021)\citenamefont {Vidmar},
  \citenamefont {Krajewski}, \citenamefont {Bonca},\ and\ \citenamefont
  {Mierzejewski}}]{Vidmar_Krajewski_Bonca_Mierzejewski_2021}%
  \BibitemOpen
  \bibfield  {author} {\bibinfo {author} {\bibfnamefont {L.}~\bibnamefont
  {Vidmar}}, \bibinfo {author} {\bibfnamefont {B.}~\bibnamefont {Krajewski}},
  \bibinfo {author} {\bibfnamefont {J.}~\bibnamefont {Bonca}}, \ and\ \bibinfo
  {author} {\bibfnamefont {M.}~\bibnamefont {Mierzejewski}},\ }\href {\doibase
  10.1103/PhysRevLett.127.230603} {\bibfield  {journal} {\bibinfo  {journal}
  {Physical Review Letters}\ }\textbf {\bibinfo {volume} {127}},\ \bibinfo
  {pages} {230603} (\bibinfo {year} {2021})}\BibitemShut {NoStop}%
\bibitem [{\citenamefont {Roy}\ and\ \citenamefont
  {Logan}(2019)}]{roy2019self}%
  \BibitemOpen
  \bibfield  {author} {\bibinfo {author} {\bibfnamefont {S.}~\bibnamefont
  {Roy}}\ and\ \bibinfo {author} {\bibfnamefont {D.~E.}\ \bibnamefont
  {Logan}},\ }\href@noop {} {\bibfield  {journal} {\bibinfo  {journal} {SciPost
  Physics}\ }\textbf {\bibinfo {volume} {7}},\ \bibinfo {pages} {042} (\bibinfo
  {year} {2019})}\BibitemShut {NoStop}%
\bibitem [{\citenamefont {Imbrie}(2016)}]{imbrie2016diagonalization}%
  \BibitemOpen
  \bibfield  {author} {\bibinfo {author} {\bibfnamefont {J.~Z.}\ \bibnamefont
  {Imbrie}},\ }\href@noop {} {\bibfield  {journal} {\bibinfo  {journal}
  {Physical review letters}\ }\textbf {\bibinfo {volume} {117}},\ \bibinfo
  {pages} {027201} (\bibinfo {year} {2016})}\BibitemShut {NoStop}%
\bibitem [{\citenamefont {Chandran}\ \emph {et~al.}(2015)\citenamefont
  {Chandran}, \citenamefont {Laumann},\ and\ \citenamefont
  {Oganesyan}}]{Chandran_Laumann_Oganesyan_2015}%
  \BibitemOpen
  \bibfield  {author} {\bibinfo {author} {\bibfnamefont {A.}~\bibnamefont
  {Chandran}}, \bibinfo {author} {\bibfnamefont {C.~R.}\ \bibnamefont
  {Laumann}}, \ and\ \bibinfo {author} {\bibfnamefont {V.}~\bibnamefont
  {Oganesyan}},\ }\href {http://arxiv.org/abs/1509.04285} {\  (\bibinfo {year}
  {2015})},\ \bibinfo {note} {arXiv:1509.04285 [cond-mat]}\BibitemShut
  {NoStop}%
\bibitem [{\citenamefont {Roy}(2022)}]{roy2022hilbert}%
  \BibitemOpen
  \bibfield  {author} {\bibinfo {author} {\bibfnamefont {S.}~\bibnamefont
  {Roy}},\ }\href@noop {} {\bibfield  {journal} {\bibinfo  {journal} {Physical
  Review B}\ }\textbf {\bibinfo {volume} {106}},\ \bibinfo {pages} {L140204}
  (\bibinfo {year} {2022})}\BibitemShut {NoStop}%
\bibitem [{\citenamefont {Roy}\ and\ \citenamefont
  {Logan}(2021{\natexlab{a}})}]{Roy_Logan_2021a}%
  \BibitemOpen
  \bibfield  {author} {\bibinfo {author} {\bibfnamefont {S.}~\bibnamefont
  {Roy}}\ and\ \bibinfo {author} {\bibfnamefont {D.~E.}\ \bibnamefont
  {Logan}},\ }\href {http://arxiv.org/abs/2106.09036} {\bibfield  {journal}
  {\bibinfo  {journal} {arXiv:2106.09036 [cond-mat, physics:quant-ph]}\ }
  (\bibinfo {year} {2021}{\natexlab{a}})},\ \bibinfo {note} {arXiv:
  2106.09036}\BibitemShut {NoStop}%
\bibitem [{\citenamefont {Macé}\ \emph {et~al.}(2019)\citenamefont {Macé},
  \citenamefont {Alet},\ and\ \citenamefont
  {Laflorencie}}]{Mace_Alet_Laflorencie_2019}%
  \BibitemOpen
  \bibfield  {author} {\bibinfo {author} {\bibfnamefont {N.}~\bibnamefont
  {Macé}}, \bibinfo {author} {\bibfnamefont {F.}~\bibnamefont {Alet}}, \ and\
  \bibinfo {author} {\bibfnamefont {N.}~\bibnamefont {Laflorencie}},\ }\href
  {\doibase 10.1103/PhysRevLett.123.180601} {\bibfield  {journal} {\bibinfo
  {journal} {Physical Review Letters}\ }\textbf {\bibinfo {volume} {123}},\
  \bibinfo {pages} {180601} (\bibinfo {year} {2019})}\BibitemShut {NoStop}%
\bibitem [{\citenamefont {Tikhonov}\ and\ \citenamefont
  {Mirlin}(2021)}]{Tikhonov_Mirlin_2021}%
  \BibitemOpen
  \bibfield  {author} {\bibinfo {author} {\bibfnamefont {K.~S.}\ \bibnamefont
  {Tikhonov}}\ and\ \bibinfo {author} {\bibfnamefont {A.~D.}\ \bibnamefont
  {Mirlin}},\ }\href {http://arxiv.org/abs/2102.05930} {\bibfield  {journal}
  {\bibinfo  {journal} {arXiv:2102.05930 [cond-mat]}\ } (\bibinfo {year}
  {2021})},\ \bibinfo {note} {arXiv: 2102.05930}\BibitemShut {NoStop}%
\bibitem [{\citenamefont {Roy}\ \emph {et~al.}(2019{\natexlab{a}})\citenamefont
  {Roy}, \citenamefont {Chalker},\ and\ \citenamefont
  {Logan}}]{Roy_Chalker_Logan_2019}%
  \BibitemOpen
  \bibfield  {author} {\bibinfo {author} {\bibfnamefont {S.}~\bibnamefont
  {Roy}}, \bibinfo {author} {\bibfnamefont {J.~T.}\ \bibnamefont {Chalker}}, \
  and\ \bibinfo {author} {\bibfnamefont {D.~E.}\ \bibnamefont {Logan}},\ }\href
  {\doibase 10.1103/PhysRevB.99.104206} {\bibfield  {journal} {\bibinfo
  {journal} {Physical Review B}\ }\textbf {\bibinfo {volume} {99}},\ \bibinfo
  {pages} {104206} (\bibinfo {year} {2019}{\natexlab{a}})},\ \bibinfo {note}
  {arXiv: 1812.06101}\BibitemShut {NoStop}%
\bibitem [{\citenamefont {Roy}\ and\ \citenamefont
  {Logan}(2021{\natexlab{b}})}]{Roy_Logan_2021b}%
  \BibitemOpen
  \bibfield  {author} {\bibinfo {author} {\bibfnamefont {S.}~\bibnamefont
  {Roy}}\ and\ \bibinfo {author} {\bibfnamefont {D.~E.}\ \bibnamefont
  {Logan}},\ }\href {\doibase 10.1103/PhysRevB.104.174201} {\bibfield
  {journal} {\bibinfo  {journal} {Physical Review B}\ }\textbf {\bibinfo
  {volume} {104}},\ \bibinfo {pages} {174201} (\bibinfo {year}
  {2021}{\natexlab{b}})}\BibitemShut {NoStop}%
\bibitem [{\citenamefont {De~Tomasi}\ \emph {et~al.}(2020)\citenamefont
  {De~Tomasi}, \citenamefont {Khaymovich}, \citenamefont {Pollmann},\ and\
  \citenamefont {Warzel}}]{Tomasi_Khaymovich_Pollmann_Warzel_2020}%
  \BibitemOpen
  \bibfield  {author} {\bibinfo {author} {\bibfnamefont {G.}~\bibnamefont
  {De~Tomasi}}, \bibinfo {author} {\bibfnamefont {I.~M.}\ \bibnamefont
  {Khaymovich}}, \bibinfo {author} {\bibfnamefont {F.}~\bibnamefont
  {Pollmann}}, \ and\ \bibinfo {author} {\bibfnamefont {S.}~\bibnamefont
  {Warzel}},\ }\href {http://arxiv.org/abs/2011.03048} {\bibfield  {journal}
  {\bibinfo  {journal} {arXiv:2011.03048 [cond-mat, physics:quant-ph]}\ }
  (\bibinfo {year} {2020})},\ \bibinfo {note} {arXiv: 2011.03048}\BibitemShut
  {NoStop}%
\bibitem [{\citenamefont {Szoldra}\ \emph {et~al.}(2021)\citenamefont
  {Szoldra}, \citenamefont {Sierant}, \citenamefont {Kottmann}, \citenamefont
  {Lewenstein},\ and\ \citenamefont
  {Zakrzewski}}]{Szoldra_Sierant_Kottmann_Lewenstein_Zakrzewski_2021}%
  \BibitemOpen
  \bibfield  {author} {\bibinfo {author} {\bibfnamefont {T.}~\bibnamefont
  {Szoldra}}, \bibinfo {author} {\bibfnamefont {P.}~\bibnamefont {Sierant}},
  \bibinfo {author} {\bibfnamefont {K.}~\bibnamefont {Kottmann}}, \bibinfo
  {author} {\bibfnamefont {M.}~\bibnamefont {Lewenstein}}, \ and\ \bibinfo
  {author} {\bibfnamefont {J.}~\bibnamefont {Zakrzewski}},\ }\href
  {http://arxiv.org/abs/2106.01811} {\bibfield  {journal} {\bibinfo  {journal}
  {arXiv:2106.01811 [cond-mat, physics:nlin, physics:physics,
  physics:quant-ph]}\ } (\bibinfo {year} {2021})},\ \bibinfo {note} {arXiv:
  2106.01811}\BibitemShut {NoStop}%
\bibitem [{\citenamefont {Morningstar}\ \emph {et~al.}(2022)\citenamefont
  {Morningstar}, \citenamefont {Colmenarez}, \citenamefont {Khemani},
  \citenamefont {Luitz},\ and\ \citenamefont
  {Huse}}]{morningstar2022avalanches}%
  \BibitemOpen
  \bibfield  {author} {\bibinfo {author} {\bibfnamefont {A.}~\bibnamefont
  {Morningstar}}, \bibinfo {author} {\bibfnamefont {L.}~\bibnamefont
  {Colmenarez}}, \bibinfo {author} {\bibfnamefont {V.}~\bibnamefont {Khemani}},
  \bibinfo {author} {\bibfnamefont {D.~J.}\ \bibnamefont {Luitz}}, \ and\
  \bibinfo {author} {\bibfnamefont {D.~A.}\ \bibnamefont {Huse}},\ }\href@noop
  {} {\bibfield  {journal} {\bibinfo  {journal} {Physical Review B}\ }\textbf
  {\bibinfo {volume} {105}},\ \bibinfo {pages} {174205} (\bibinfo {year}
  {2022})}\BibitemShut {NoStop}%
\bibitem [{\citenamefont {Roy}\ \emph {et~al.}(2019{\natexlab{b}})\citenamefont
  {Roy}, \citenamefont {Logan},\ and\ \citenamefont
  {Chalker}}]{Roy_Logan_Chalker_2019}%
  \BibitemOpen
  \bibfield  {author} {\bibinfo {author} {\bibfnamefont {S.}~\bibnamefont
  {Roy}}, \bibinfo {author} {\bibfnamefont {D.~E.}\ \bibnamefont {Logan}}, \
  and\ \bibinfo {author} {\bibfnamefont {J.~T.}\ \bibnamefont {Chalker}},\
  }\href {\doibase 10.1103/PhysRevB.99.220201} {\bibfield  {journal} {\bibinfo
  {journal} {Physical Review B}\ }\textbf {\bibinfo {volume} {99}},\ \bibinfo
  {pages} {220201} (\bibinfo {year} {2019}{\natexlab{b}})}\BibitemShut
  {NoStop}%
\bibitem [{\citenamefont {Goblot}\ \emph {et~al.}(2020)\citenamefont {Goblot},
  \citenamefont {Štrkalj}, \citenamefont {Pernet}, \citenamefont {Lado},
  \citenamefont {Dorow}, \citenamefont {Lemaître}, \citenamefont {Gratiet},
  \citenamefont {Harouri}, \citenamefont {Sagnes}, \citenamefont {Ravets},
  \citenamefont {Amo}, \citenamefont {Bloch},\ and\ \citenamefont
  {Zilberberg}}]{Goblot_Strkalj_Pernet_Lado_Dorow_Lemaitre_Gratiet_Harouri_Sagnes_Ravets}%
  \BibitemOpen
  \bibfield  {author} {\bibinfo {author} {\bibfnamefont {V.}~\bibnamefont
  {Goblot}}, \bibinfo {author} {\bibfnamefont {A.}~\bibnamefont {Štrkalj}},
  \bibinfo {author} {\bibfnamefont {N.}~\bibnamefont {Pernet}}, \bibinfo
  {author} {\bibfnamefont {J.~L.}\ \bibnamefont {Lado}}, \bibinfo {author}
  {\bibfnamefont {C.}~\bibnamefont {Dorow}}, \bibinfo {author} {\bibfnamefont
  {A.}~\bibnamefont {Lemaître}}, \bibinfo {author} {\bibfnamefont {L.~L.}\
  \bibnamefont {Gratiet}}, \bibinfo {author} {\bibfnamefont {A.}~\bibnamefont
  {Harouri}}, \bibinfo {author} {\bibfnamefont {I.}~\bibnamefont {Sagnes}},
  \bibinfo {author} {\bibfnamefont {S.}~\bibnamefont {Ravets}}, \bibinfo
  {author} {\bibfnamefont {A.}~\bibnamefont {Amo}}, \bibinfo {author}
  {\bibfnamefont {J.}~\bibnamefont {Bloch}}, \ and\ \bibinfo {author}
  {\bibfnamefont {O.}~\bibnamefont {Zilberberg}},\ }\href {\doibase
  10.1038/s41567-020-0908-7} {\bibfield  {journal} {\bibinfo  {journal} {Nature
  Physics}\ }\textbf {\bibinfo {volume} {16}},\ \bibinfo {pages} {832–836}
  (\bibinfo {year} {2020})},\ \bibinfo {note} {arXiv: 1911.07809}\BibitemShut
  {NoStop}%
\bibitem [{\citenamefont {Logan}\ and\ \citenamefont
  {Welsh}(2019)}]{Logan_Welsh_2019}%
  \BibitemOpen
  \bibfield  {author} {\bibinfo {author} {\bibfnamefont {D.~E.}\ \bibnamefont
  {Logan}}\ and\ \bibinfo {author} {\bibfnamefont {S.}~\bibnamefont {Welsh}},\
  }\href {\doibase 10.1103/PhysRevB.99.045131} {\bibfield  {journal} {\bibinfo
  {journal} {Physical Review B}\ }\textbf {\bibinfo {volume} {99}},\ \bibinfo
  {pages} {045131} (\bibinfo {year} {2019})},\ \bibinfo {note} {arXiv:
  1806.01688}\BibitemShut {NoStop}%
\bibitem [{\citenamefont {Thiery}\ \emph {et~al.}(2018)\citenamefont {Thiery},
  \citenamefont {Huveneers}, \citenamefont {Müller},\ and\ \citenamefont
  {De~Roeck}}]{Thiery_Huveneers_Muller_De}%
  \BibitemOpen
  \bibfield  {author} {\bibinfo {author} {\bibfnamefont {T.}~\bibnamefont
  {Thiery}}, \bibinfo {author} {\bibfnamefont {F.}~\bibnamefont {Huveneers}},
  \bibinfo {author} {\bibfnamefont {M.}~\bibnamefont {Müller}}, \ and\
  \bibinfo {author} {\bibfnamefont {W.}~\bibnamefont {De~Roeck}},\ }\href
  {\doibase 10.1103/PhysRevLett.121.140601} {\bibfield  {journal} {\bibinfo
  {journal} {Physical Review Letters}\ }\textbf {\bibinfo {volume} {121}},\
  \bibinfo {pages} {140601} (\bibinfo {year} {2018})},\ \bibinfo {note} {arXiv:
  1706.09338}\BibitemShut {NoStop}%
\bibitem [{\citenamefont {Prelovšek}\ \emph {et~al.}(2018)\citenamefont
  {Prelovšek}, \citenamefont {Barišić},\ and\ \citenamefont
  {Mierzejewski}}]{Prelovsek_Barisic_Mierzejewski_2018}%
  \BibitemOpen
  \bibfield  {author} {\bibinfo {author} {\bibfnamefont {P.}~\bibnamefont
  {Prelovšek}}, \bibinfo {author} {\bibfnamefont {O.~S.}\ \bibnamefont
  {Barišić}}, \ and\ \bibinfo {author} {\bibfnamefont {M.}~\bibnamefont
  {Mierzejewski}},\ }\href {\doibase 10.1103/PhysRevB.97.035104} {\bibfield
  {journal} {\bibinfo  {journal} {Physical Review B}\ }\textbf {\bibinfo
  {volume} {97}},\ \bibinfo {pages} {035104} (\bibinfo {year}
  {2018})}\BibitemShut {NoStop}%
\bibitem [{\citenamefont {Filoche}\ and\ \citenamefont
  {Mayboroda}(2012)}]{filoche2012universal}%
  \BibitemOpen
  \bibfield  {author} {\bibinfo {author} {\bibfnamefont {M.}~\bibnamefont
  {Filoche}}\ and\ \bibinfo {author} {\bibfnamefont {S.}~\bibnamefont
  {Mayboroda}},\ }\href@noop {} {\bibfield  {journal} {\bibinfo  {journal}
  {Proceedings of the National Academy of Sciences}\ }\textbf {\bibinfo
  {volume} {109}},\ \bibinfo {pages} {14761} (\bibinfo {year}
  {2012})}\BibitemShut {NoStop}%
\bibitem [{\citenamefont {Arnold}\ \emph {et~al.}(2018)\citenamefont {Arnold},
  \citenamefont {David}, \citenamefont {Filoche}, \citenamefont {Jerison},\
  and\ \citenamefont
  {Mayboroda}}]{Arnold_David_Filoche_Jerison_Mayboroda_2018}%
  \BibitemOpen
  \bibfield  {author} {\bibinfo {author} {\bibfnamefont {D.}~\bibnamefont
  {Arnold}}, \bibinfo {author} {\bibfnamefont {G.}~\bibnamefont {David}},
  \bibinfo {author} {\bibfnamefont {M.}~\bibnamefont {Filoche}}, \bibinfo
  {author} {\bibfnamefont {D.}~\bibnamefont {Jerison}}, \ and\ \bibinfo
  {author} {\bibfnamefont {S.}~\bibnamefont {Mayboroda}},\ }\href
  {http://arxiv.org/abs/1711.04888} {\bibfield  {journal} {\bibinfo  {journal}
  {arXiv:1711.04888 [math]}\ } (\bibinfo {year} {2018})},\ \bibinfo {note}
  {arXiv: 1711.04888}\BibitemShut {NoStop}%
\bibitem [{\citenamefont {Arnold}\ \emph {et~al.}(2015)\citenamefont {Arnold},
  \citenamefont {David}, \citenamefont {Jerison}, \citenamefont {Mayboroda},\
  and\ \citenamefont {Filoche}}]{Arnold_David_Jerison_Mayboroda_Filoche_2015}%
  \BibitemOpen
  \bibfield  {author} {\bibinfo {author} {\bibfnamefont {D.~L.}\ \bibnamefont
  {Arnold}}, \bibinfo {author} {\bibfnamefont {G.}~\bibnamefont {David}},
  \bibinfo {author} {\bibfnamefont {D.}~\bibnamefont {Jerison}}, \bibinfo
  {author} {\bibfnamefont {S.}~\bibnamefont {Mayboroda}}, \ and\ \bibinfo
  {author} {\bibfnamefont {M.}~\bibnamefont {Filoche}},\ }\href
  {https://hal.archives-ouvertes.fr/hal-01150766} {\enquote {\bibinfo {title}
  {The effective confining potential of quantum states in disordered media},}\
  } (\bibinfo {year} {2015})\BibitemShut {NoStop}%
\bibitem [{\citenamefont {Pelletier}\ \emph {et~al.}(2021)\citenamefont
  {Pelletier}, \citenamefont {Delande}, \citenamefont {Josse}, \citenamefont
  {Aspect}, \citenamefont {Mayboroda}, \citenamefont {Arnold},\ and\
  \citenamefont
  {Filoche}}]{Pelletier_Delande_Josse_Aspect_Mayboroda_Arnold_Filoche_2021}%
  \BibitemOpen
  \bibfield  {author} {\bibinfo {author} {\bibfnamefont {P.}~\bibnamefont
  {Pelletier}}, \bibinfo {author} {\bibfnamefont {D.}~\bibnamefont {Delande}},
  \bibinfo {author} {\bibfnamefont {V.}~\bibnamefont {Josse}}, \bibinfo
  {author} {\bibfnamefont {A.}~\bibnamefont {Aspect}}, \bibinfo {author}
  {\bibfnamefont {S.}~\bibnamefont {Mayboroda}}, \bibinfo {author}
  {\bibfnamefont {D.}~\bibnamefont {Arnold}}, \ and\ \bibinfo {author}
  {\bibfnamefont {M.}~\bibnamefont {Filoche}},\ }\href
  {http://arxiv.org/abs/2111.13155} {\bibfield  {journal} {\bibinfo  {journal}
  {arXiv:2111.13155 [quant-ph]}\ } (\bibinfo {year} {2021})},\ \bibinfo {note}
  {arXiv: 2111.13155}\BibitemShut {NoStop}%
\bibitem [{\citenamefont {Shamailov}\ \emph {et~al.}(2021)\citenamefont
  {Shamailov}, \citenamefont {Brown}, \citenamefont {Haase},\ and\
  \citenamefont {Hoogerland}}]{Shamailov_Brown_Haase_Hoogerland_2021}%
  \BibitemOpen
  \bibfield  {author} {\bibinfo {author} {\bibfnamefont {S.}~\bibnamefont
  {Shamailov}}, \bibinfo {author} {\bibfnamefont {D.~J.}\ \bibnamefont
  {Brown}}, \bibinfo {author} {\bibfnamefont {T.~A.}\ \bibnamefont {Haase}}, \
  and\ \bibinfo {author} {\bibfnamefont {M.}~\bibnamefont {Hoogerland}},\
  }\href {\doibase 10.21468/SciPostPhysCore.4.2.017} {\bibfield  {journal}
  {\bibinfo  {journal} {SciPost Physics Core}\ }\textbf {\bibinfo {volume}
  {4}},\ \bibinfo {pages} {017} (\bibinfo {year} {2021})}\BibitemShut {NoStop}%
\bibitem [{\citenamefont {Li}\ \emph {et~al.}(2017)\citenamefont {Li},
  \citenamefont {Piccardo}, \citenamefont {Lu}, \citenamefont {Mayboroda},
  \citenamefont {Martinelli}, \citenamefont {Peretti}, \citenamefont {Speck},
  \citenamefont {Weisbuch}, \citenamefont {Filoche},\ and\ \citenamefont
  {Wu}}]{li2017localization}%
  \BibitemOpen
  \bibfield  {author} {\bibinfo {author} {\bibfnamefont {C.-K.}\ \bibnamefont
  {Li}}, \bibinfo {author} {\bibfnamefont {M.}~\bibnamefont {Piccardo}},
  \bibinfo {author} {\bibfnamefont {L.-S.}\ \bibnamefont {Lu}}, \bibinfo
  {author} {\bibfnamefont {S.}~\bibnamefont {Mayboroda}}, \bibinfo {author}
  {\bibfnamefont {L.}~\bibnamefont {Martinelli}}, \bibinfo {author}
  {\bibfnamefont {J.}~\bibnamefont {Peretti}}, \bibinfo {author} {\bibfnamefont
  {J.~S.}\ \bibnamefont {Speck}}, \bibinfo {author} {\bibfnamefont
  {C.}~\bibnamefont {Weisbuch}}, \bibinfo {author} {\bibfnamefont
  {M.}~\bibnamefont {Filoche}}, \ and\ \bibinfo {author} {\bibfnamefont
  {Y.-R.}\ \bibnamefont {Wu}},\ }\href@noop {} {\bibfield  {journal} {\bibinfo
  {journal} {Physical Review B}\ }\textbf {\bibinfo {volume} {95}},\ \bibinfo
  {pages} {144206} (\bibinfo {year} {2017})}\BibitemShut {NoStop}%
\bibitem [{\citenamefont {Kirkpatrick}(1973)}]{kirkpatrick1973percolation}%
  \BibitemOpen
  \bibfield  {author} {\bibinfo {author} {\bibfnamefont {S.}~\bibnamefont
  {Kirkpatrick}},\ }\href@noop {} {\bibfield  {journal} {\bibinfo  {journal}
  {Reviews of modern physics}\ }\textbf {\bibinfo {volume} {45}},\ \bibinfo
  {pages} {574} (\bibinfo {year} {1973})}\BibitemShut {NoStop}%
\bibitem [{\citenamefont {Balasubramanian}\ \emph {et~al.}(2020)\citenamefont
  {Balasubramanian}, \citenamefont {Liao},\ and\ \citenamefont
  {Galitski}}]{Balasubramanian_Liao_Galitski_2020}%
  \BibitemOpen
  \bibfield  {author} {\bibinfo {author} {\bibfnamefont {S.}~\bibnamefont
  {Balasubramanian}}, \bibinfo {author} {\bibfnamefont {Y.}~\bibnamefont
  {Liao}}, \ and\ \bibinfo {author} {\bibfnamefont {V.}~\bibnamefont
  {Galitski}},\ }\href {\doibase 10.1103/PhysRevB.101.014201} {\bibfield
  {journal} {\bibinfo  {journal} {Physical Review B}\ }\textbf {\bibinfo
  {volume} {101}},\ \bibinfo {pages} {014201} (\bibinfo {year}
  {2020})}\BibitemShut {NoStop}%
\bibitem [{\citenamefont {Bel-Hadj-Aissa}\ \emph {et~al.}(2021)\citenamefont
  {Bel-Hadj-Aissa}, \citenamefont {Gori}, \citenamefont {Franzosi},\ and\
  \citenamefont {Pettini}}]{Bel-Hadj-Aissa_Gori_Franzosi_Pettini_2021}%
  \BibitemOpen
  \bibfield  {author} {\bibinfo {author} {\bibfnamefont {G.}~\bibnamefont
  {Bel-Hadj-Aissa}}, \bibinfo {author} {\bibfnamefont {M.}~\bibnamefont
  {Gori}}, \bibinfo {author} {\bibfnamefont {R.}~\bibnamefont {Franzosi}}, \
  and\ \bibinfo {author} {\bibfnamefont {M.}~\bibnamefont {Pettini}},\ }\href
  {\doibase 10.1088/1742-5468/abda27} {\bibfield  {journal} {\bibinfo
  {journal} {Journal of Statistical Mechanics: Theory and Experiment}\ }\textbf
  {\bibinfo {volume} {2021}},\ \bibinfo {pages} {023206} (\bibinfo {year}
  {2021})}\BibitemShut {NoStop}%
\bibitem [{\citenamefont {Gori}\ \emph {et~al.}(2022)\citenamefont {Gori},
  \citenamefont {Franzosi}, \citenamefont {Pettini},\ and\ \citenamefont
  {Pettini}}]{Gori_Franzosi_Pettini_Pettini_2022}%
  \BibitemOpen
  \bibfield  {author} {\bibinfo {author} {\bibfnamefont {M.}~\bibnamefont
  {Gori}}, \bibinfo {author} {\bibfnamefont {R.}~\bibnamefont {Franzosi}},
  \bibinfo {author} {\bibfnamefont {G.}~\bibnamefont {Pettini}}, \ and\
  \bibinfo {author} {\bibfnamefont {M.}~\bibnamefont {Pettini}},\ }\href
  {\doibase 10.1088/1751-8121/ac7f09} {\bibfield  {journal} {\bibinfo
  {journal} {Journal of Physics A: Mathematical and Theoretical}\ } (\bibinfo
  {year} {2022}),\ 10.1088/1751-8121/ac7f09},\ \bibinfo {note}
  {arXiv:2207.05077 [cond-mat, physics:math-ph, physics:physics]}\BibitemShut
  {NoStop}%
\bibitem [{\citenamefont {Sale}\ \emph {et~al.}(2022)\citenamefont {Sale},
  \citenamefont {Giansiracusa},\ and\ \citenamefont
  {Lucini}}]{Sale_Giansiracusa_Lucini_2022}%
  \BibitemOpen
  \bibfield  {author} {\bibinfo {author} {\bibfnamefont {N.}~\bibnamefont
  {Sale}}, \bibinfo {author} {\bibfnamefont {J.}~\bibnamefont {Giansiracusa}},
  \ and\ \bibinfo {author} {\bibfnamefont {B.}~\bibnamefont {Lucini}},\ }\href
  {\doibase 10.1103/PhysRevE.105.024121} {\bibfield  {journal} {\bibinfo
  {journal} {Physical Review E}\ }\textbf {\bibinfo {volume} {105}},\ \bibinfo
  {pages} {024121} (\bibinfo {year} {2022})}\BibitemShut {NoStop}%
\bibitem [{\citenamefont {Donato}\ \emph {et~al.}(2016)\citenamefont {Donato},
  \citenamefont {Gori}, \citenamefont {Pettini}, \citenamefont {Petri},
  \citenamefont {De~Nigris}, \citenamefont {Franzosi},\ and\ \citenamefont
  {Vaccarino}}]{donato2016persistent}%
  \BibitemOpen
  \bibfield  {author} {\bibinfo {author} {\bibfnamefont {I.}~\bibnamefont
  {Donato}}, \bibinfo {author} {\bibfnamefont {M.}~\bibnamefont {Gori}},
  \bibinfo {author} {\bibfnamefont {M.}~\bibnamefont {Pettini}}, \bibinfo
  {author} {\bibfnamefont {G.}~\bibnamefont {Petri}}, \bibinfo {author}
  {\bibfnamefont {S.}~\bibnamefont {De~Nigris}}, \bibinfo {author}
  {\bibfnamefont {R.}~\bibnamefont {Franzosi}}, \ and\ \bibinfo {author}
  {\bibfnamefont {F.}~\bibnamefont {Vaccarino}},\ }\href@noop {} {\bibfield
  {journal} {\bibinfo  {journal} {Physical Review E}\ }\textbf {\bibinfo
  {volume} {93}},\ \bibinfo {pages} {052138} (\bibinfo {year}
  {2016})}\BibitemShut {NoStop}%
\bibitem [{\citenamefont {Herviou}\ and\ \citenamefont
  {Bardarson}(2020)}]{Herviou_Bardarson_2020}%
  \BibitemOpen
  \bibfield  {author} {\bibinfo {author} {\bibfnamefont {L.}~\bibnamefont
  {Herviou}}\ and\ \bibinfo {author} {\bibfnamefont {J.~H.}\ \bibnamefont
  {Bardarson}},\ }\href {\doibase 10.1103/PhysRevB.101.220201} {\bibfield
  {journal} {\bibinfo  {journal} {Physical Review B}\ }\textbf {\bibinfo
  {volume} {101}},\ \bibinfo {pages} {220201} (\bibinfo {year} {2020})},\
  \bibinfo {note} {arXiv: 2004.02903}\BibitemShut {NoStop}%
\bibitem [{\citenamefont {Lemut}\ \emph {et~al.}(2020)\citenamefont {Lemut},
  \citenamefont {Pacholski}, \citenamefont {Ovdat}, \citenamefont {Grabsch},
  \citenamefont {Tworzyd{\l}o},\ and\ \citenamefont
  {Beenakker}}]{lemut2020localization}%
  \BibitemOpen
  \bibfield  {author} {\bibinfo {author} {\bibfnamefont {G.}~\bibnamefont
  {Lemut}}, \bibinfo {author} {\bibfnamefont {M.}~\bibnamefont {Pacholski}},
  \bibinfo {author} {\bibfnamefont {O.}~\bibnamefont {Ovdat}}, \bibinfo
  {author} {\bibfnamefont {A.}~\bibnamefont {Grabsch}}, \bibinfo {author}
  {\bibfnamefont {J.}~\bibnamefont {Tworzyd{\l}o}}, \ and\ \bibinfo {author}
  {\bibfnamefont {C.}~\bibnamefont {Beenakker}},\ }\href@noop {} {\bibfield
  {journal} {\bibinfo  {journal} {Physical Review B}\ }\textbf {\bibinfo
  {volume} {101}},\ \bibinfo {pages} {081405} (\bibinfo {year}
  {2020})}\BibitemShut {NoStop}%
\bibitem [{\citenamefont {Zomorodian}\ and\ \citenamefont
  {Carlsson}(2005)}]{Zomorodian_Carlsson_2005}%
  \BibitemOpen
  \bibfield  {author} {\bibinfo {author} {\bibfnamefont {A.}~\bibnamefont
  {Zomorodian}}\ and\ \bibinfo {author} {\bibfnamefont {G.}~\bibnamefont
  {Carlsson}},\ }\href {\doibase 10.1007/s00454-004-1146-y} {\bibfield
  {journal} {\bibinfo  {journal} {Discrete I\& Computational Geometry}\
  }\textbf {\bibinfo {volume} {33}},\ \bibinfo {pages} {249–274} (\bibinfo
  {year} {2005})}\BibitemShut {NoStop}%
\bibitem [{\citenamefont {He}\ \emph {et~al.}(2022)\citenamefont {He},
  \citenamefont {Xia}, \citenamefont {Angelakis}, \citenamefont {Song},
  \citenamefont {Chen},\ and\ \citenamefont
  {Leykam}}]{He_Xia_Angelakis_Song_Chen_Leykam_2022}%
  \BibitemOpen
  \bibfield  {author} {\bibinfo {author} {\bibfnamefont {Y.}~\bibnamefont
  {He}}, \bibinfo {author} {\bibfnamefont {S.}~\bibnamefont {Xia}}, \bibinfo
  {author} {\bibfnamefont {D.~G.}\ \bibnamefont {Angelakis}}, \bibinfo {author}
  {\bibfnamefont {D.}~\bibnamefont {Song}}, \bibinfo {author} {\bibfnamefont
  {Z.}~\bibnamefont {Chen}}, \ and\ \bibinfo {author} {\bibfnamefont
  {D.}~\bibnamefont {Leykam}},\ }\href {http://arxiv.org/abs/2204.13276}
  {\bibfield  {journal} {\bibinfo  {journal} {arXiv:2204.13276 [cond-mat,
  physics:physics, physics:quant-ph]}\ } (\bibinfo {year} {2022})},\ \bibinfo
  {note} {arXiv: 2204.13276}\BibitemShut {NoStop}%
\bibitem [{\citenamefont {Cole}\ \emph {et~al.}(2021)\citenamefont {Cole},
  \citenamefont {Loges},\ and\ \citenamefont {Shiu}}]{Cole_Loges_Shiu_2021}%
  \BibitemOpen
  \bibfield  {author} {\bibinfo {author} {\bibfnamefont {A.}~\bibnamefont
  {Cole}}, \bibinfo {author} {\bibfnamefont {G.~J.}\ \bibnamefont {Loges}}, \
  and\ \bibinfo {author} {\bibfnamefont {G.}~\bibnamefont {Shiu}},\ }\href
  {\doibase 10.1103/PhysRevB.104.104426} {\bibfield  {journal} {\bibinfo
  {journal} {Physical Review B}\ }\textbf {\bibinfo {volume} {104}},\ \bibinfo
  {pages} {104426} (\bibinfo {year} {2021})}\BibitemShut {NoStop}%
\bibitem [{\citenamefont {Olsthoorn}\ and\ \citenamefont
  {Balatsky}(2021)}]{Olsthoorn_Balatsky_2021}%
  \BibitemOpen
  \bibfield  {author} {\bibinfo {author} {\bibfnamefont {B.}~\bibnamefont
  {Olsthoorn}}\ and\ \bibinfo {author} {\bibfnamefont {A.~V.}\ \bibnamefont
  {Balatsky}},\ }\href {http://arxiv.org/abs/2110.10214} {\bibfield  {journal}
  {\bibinfo  {journal} {arXiv:2110.10214 [cond-mat, physics:quant-ph]}\ }
  (\bibinfo {year} {2021})},\ \bibinfo {note} {arXiv: 2110.10214}\BibitemShut
  {NoStop}%
\bibitem [{\citenamefont {Filoche}\ \emph {et~al.}(2021)\citenamefont
  {Filoche}, \citenamefont {Mayboroda},\ and\ \citenamefont
  {Tao}}]{Filoche_Mayboroda_Tao_2021}%
  \BibitemOpen
  \bibfield  {author} {\bibinfo {author} {\bibfnamefont {M.}~\bibnamefont
  {Filoche}}, \bibinfo {author} {\bibfnamefont {S.}~\bibnamefont {Mayboroda}},
  \ and\ \bibinfo {author} {\bibfnamefont {T.}~\bibnamefont {Tao}},\ }\href
  {\doibase 10.1063/5.0042629} {\bibfield  {journal} {\bibinfo  {journal}
  {Journal of Mathematical Physics}\ }\textbf {\bibinfo {volume} {62}},\
  \bibinfo {pages} {041902} (\bibinfo {year} {2021})}\BibitemShut {NoStop}%
\bibitem [{\citenamefont {Lyra}\ \emph {et~al.}(2015)\citenamefont {Lyra},
  \citenamefont {Mayboroda},\ and\ \citenamefont {Filoche}}]{lyra2015dual}%
  \BibitemOpen
  \bibfield  {author} {\bibinfo {author} {\bibfnamefont {M.~L.}\ \bibnamefont
  {Lyra}}, \bibinfo {author} {\bibfnamefont {S.}~\bibnamefont {Mayboroda}}, \
  and\ \bibinfo {author} {\bibfnamefont {M.}~\bibnamefont {Filoche}},\
  }\href@noop {} {\bibfield  {journal} {\bibinfo  {journal} {EPL (Europhysics
  Letters)}\ }\textbf {\bibinfo {volume} {109}},\ \bibinfo {pages} {47001}
  (\bibinfo {year} {2015})}\BibitemShut {NoStop}%
\bibitem [{\citenamefont {Murphy}\ \emph {et~al.}(2011)\citenamefont {Murphy},
  \citenamefont {Wortis},\ and\ \citenamefont
  {Atkinson}}]{Murphy_Wortis_Atkinson_2011}%
  \BibitemOpen
  \bibfield  {author} {\bibinfo {author} {\bibfnamefont {N.~C.}\ \bibnamefont
  {Murphy}}, \bibinfo {author} {\bibfnamefont {R.}~\bibnamefont {Wortis}}, \
  and\ \bibinfo {author} {\bibfnamefont {W.~A.}\ \bibnamefont {Atkinson}},\
  }\href {\doibase 10.1103/PhysRevB.83.184206} {\bibfield  {journal} {\bibinfo
  {journal} {Physical Review B}\ }\textbf {\bibinfo {volume} {83}},\ \bibinfo
  {pages} {184206} (\bibinfo {year} {2011})},\ \bibinfo {note} {arXiv:1011.0659
  [cond-mat]}\BibitemShut {NoStop}%
\bibitem [{\citenamefont {Otter}\ \emph {et~al.}(2017)\citenamefont {Otter},
  \citenamefont {Porter}, \citenamefont {Tillmann}, \citenamefont {Grindrod},\
  and\ \citenamefont {Harrington}}]{otter2017roadmap}%
  \BibitemOpen
  \bibfield  {author} {\bibinfo {author} {\bibfnamefont {N.}~\bibnamefont
  {Otter}}, \bibinfo {author} {\bibfnamefont {M.~A.}\ \bibnamefont {Porter}},
  \bibinfo {author} {\bibfnamefont {U.}~\bibnamefont {Tillmann}}, \bibinfo
  {author} {\bibfnamefont {P.}~\bibnamefont {Grindrod}}, \ and\ \bibinfo
  {author} {\bibfnamefont {H.~A.}\ \bibnamefont {Harrington}},\ }\href@noop {}
  {\bibfield  {journal} {\bibinfo  {journal} {EPJ Data Science}\ }\textbf
  {\bibinfo {volume} {6}},\ \bibinfo {pages} {1} (\bibinfo {year}
  {2017})}\BibitemShut {NoStop}%
\bibitem [{\citenamefont {Cohen-Steiner}\ \emph {et~al.}(2007)\citenamefont
  {Cohen-Steiner}, \citenamefont {Edelsbrunner},\ and\ \citenamefont
  {Harer}}]{Cohen-Steiner_Edelsbrunner_Harer_2007}%
  \BibitemOpen
  \bibfield  {author} {\bibinfo {author} {\bibfnamefont {D.}~\bibnamefont
  {Cohen-Steiner}}, \bibinfo {author} {\bibfnamefont {H.}~\bibnamefont
  {Edelsbrunner}}, \ and\ \bibinfo {author} {\bibfnamefont {J.}~\bibnamefont
  {Harer}},\ }\href {\doibase 10.1007/s00454-006-1276-5} {\bibfield  {journal}
  {\bibinfo  {journal} {Discrete I\& Computational Geometry}\ }\textbf
  {\bibinfo {volume} {37}},\ \bibinfo {pages} {103–120} (\bibinfo {year}
  {2007})}\BibitemShut {NoStop}%
\bibitem [{\citenamefont {Bauer}\ and\ \citenamefont
  {Lesnick}(2015)}]{Bauer_Lesnick_2015}%
  \BibitemOpen
  \bibfield  {author} {\bibinfo {author} {\bibfnamefont {U.}~\bibnamefont
  {Bauer}}\ and\ \bibinfo {author} {\bibfnamefont {M.}~\bibnamefont
  {Lesnick}},\ }\href {\doibase 10.20382/jocg.v6i2a9} {\bibfield  {journal}
  {\bibinfo  {journal} {Journal of Computational Geometry}\ ,\ \bibinfo {pages}
  {162}} (\bibinfo {year} {2015})},\ \bibinfo {note} {arXiv:
  1311.3681}\BibitemShut {NoStop}%
\bibitem [{\citenamefont {Carlsson}\ \emph {et~al.}(2009)\citenamefont
  {Carlsson}, \citenamefont {De~Silva},\ and\ \citenamefont
  {Morozov}}]{carlsson2009zigzag}%
  \BibitemOpen
  \bibfield  {author} {\bibinfo {author} {\bibfnamefont {G.}~\bibnamefont
  {Carlsson}}, \bibinfo {author} {\bibfnamefont {V.}~\bibnamefont {De~Silva}},
  \ and\ \bibinfo {author} {\bibfnamefont {D.}~\bibnamefont {Morozov}},\ }in\
  \href@noop {} {\emph {\bibinfo {booktitle} {Proceedings of the twenty-fifth
  annual symposium on Computational geometry}}}\ (\bibinfo {year} {2009})\ pp.\
  \bibinfo {pages} {247--256}\BibitemShut {NoStop}%
\bibitem [{\citenamefont {Devakul}\ and\ \citenamefont
  {Singh}(2015)}]{Devakul_Singh_2015}%
  \BibitemOpen
  \bibfield  {author} {\bibinfo {author} {\bibfnamefont {T.}~\bibnamefont
  {Devakul}}\ and\ \bibinfo {author} {\bibfnamefont {R.~R.}\ \bibnamefont
  {Singh}},\ }\href {\doibase 10.1103/PhysRevLett.115.187201} {\bibfield
  {journal} {\bibinfo  {journal} {Physical Review Letters}\ }\textbf {\bibinfo
  {volume} {115}},\ \bibinfo {pages} {187201} (\bibinfo {year}
  {2015})}\BibitemShut {NoStop}%
\bibitem [{\citenamefont {Doggen}\ \emph {et~al.}(2018)\citenamefont {Doggen},
  \citenamefont {Schindler}, \citenamefont {Tikhonov}, \citenamefont {Mirlin},
  \citenamefont {Neupert}, \citenamefont {Polyakov},\ and\ \citenamefont
  {Gornyi}}]{Doggen_Schindler_Tikhonov_Mirlin_Neupert_Polyakov_Gornyi_2018}%
  \BibitemOpen
  \bibfield  {author} {\bibinfo {author} {\bibfnamefont {E.~V.~H.}\
  \bibnamefont {Doggen}}, \bibinfo {author} {\bibfnamefont {F.}~\bibnamefont
  {Schindler}}, \bibinfo {author} {\bibfnamefont {K.~S.}\ \bibnamefont
  {Tikhonov}}, \bibinfo {author} {\bibfnamefont {A.~D.}\ \bibnamefont
  {Mirlin}}, \bibinfo {author} {\bibfnamefont {T.}~\bibnamefont {Neupert}},
  \bibinfo {author} {\bibfnamefont {D.~G.}\ \bibnamefont {Polyakov}}, \ and\
  \bibinfo {author} {\bibfnamefont {I.~V.}\ \bibnamefont {Gornyi}},\ }\href
  {\doibase 10.1103/PhysRevB.98.174202} {\bibfield  {journal} {\bibinfo
  {journal} {Physical Review B}\ }\textbf {\bibinfo {volume} {98}},\ \bibinfo
  {pages} {174202} (\bibinfo {year} {2018})}\BibitemShut {NoStop}%
\bibitem [{\citenamefont {Serbyn}\ and\ \citenamefont
  {Moore}(2016)}]{Serbyn_Moore_2016}%
  \BibitemOpen
  \bibfield  {author} {\bibinfo {author} {\bibfnamefont {M.}~\bibnamefont
  {Serbyn}}\ and\ \bibinfo {author} {\bibfnamefont {J.~E.}\ \bibnamefont
  {Moore}},\ }\href {\doibase 10.1103/PhysRevB.93.041424} {\bibfield  {journal}
  {\bibinfo  {journal} {Physical Review B}\ }\textbf {\bibinfo {volume} {93}},\
  \bibinfo {pages} {041424} (\bibinfo {year} {2016})},\ \bibinfo {note} {arXiv:
  1508.07293}\BibitemShut {NoStop}%
\bibitem [{\citenamefont {Sierant}\ and\ \citenamefont
  {Zakrzewski}(2019)}]{Sierant_Zakrzewski_2019}%
  \BibitemOpen
  \bibfield  {author} {\bibinfo {author} {\bibfnamefont {P.}~\bibnamefont
  {Sierant}}\ and\ \bibinfo {author} {\bibfnamefont {J.}~\bibnamefont
  {Zakrzewski}},\ }\href {\doibase 10.1103/PhysRevB.99.104205} {\bibfield
  {journal} {\bibinfo  {journal} {Physical Review B}\ }\textbf {\bibinfo
  {volume} {99}},\ \bibinfo {pages} {104205} (\bibinfo {year} {2019})},\
  \bibinfo {note} {arXiv: 1808.02795}\BibitemShut {NoStop}%
\bibitem [{\citenamefont {Jaquette}\ and\ \citenamefont
  {Schweinhart}(2020)}]{Jaquette_Schweinhart_2020}%
  \BibitemOpen
  \bibfield  {author} {\bibinfo {author} {\bibfnamefont {J.}~\bibnamefont
  {Jaquette}}\ and\ \bibinfo {author} {\bibfnamefont {B.}~\bibnamefont
  {Schweinhart}},\ }\href {\doibase 10.1016/j.cnsns.2019.105163} {\bibfield
  {journal} {\bibinfo  {journal} {Communications in Nonlinear Science and
  Numerical Simulation}\ }\textbf {\bibinfo {volume} {84}},\ \bibinfo {pages}
  {105163} (\bibinfo {year} {2020})}\BibitemShut {NoStop}%
\bibitem [{\citenamefont {Bubenik}\ \emph {et~al.}(2020)\citenamefont
  {Bubenik}, \citenamefont {Hull}, \citenamefont {Patel},\ and\ \citenamefont
  {Whittle}}]{Bubenik_Hull_Patel_Whittle_2020}%
  \BibitemOpen
  \bibfield  {author} {\bibinfo {author} {\bibfnamefont {P.}~\bibnamefont
  {Bubenik}}, \bibinfo {author} {\bibfnamefont {M.}~\bibnamefont {Hull}},
  \bibinfo {author} {\bibfnamefont {D.}~\bibnamefont {Patel}}, \ and\ \bibinfo
  {author} {\bibfnamefont {B.}~\bibnamefont {Whittle}},\ }\href {\doibase
  10.1088/1361-6420/ab4ac0} {\bibfield  {journal} {\bibinfo  {journal} {Inverse
  Problems}\ }\textbf {\bibinfo {volume} {36}},\ \bibinfo {pages} {025008}
  (\bibinfo {year} {2020})},\ \bibinfo {note} {arXiv: 1905.13196}\BibitemShut
  {NoStop}%
\bibitem [{\citenamefont {Bollob{\'a}s}(1998)}]{bollobas1998random}%
  \BibitemOpen
  \bibfield  {author} {\bibinfo {author} {\bibfnamefont {B.}~\bibnamefont
  {Bollob{\'a}s}},\ }in\ \href@noop {} {\emph {\bibinfo {booktitle} {Modern
  graph theory}}}\ (\bibinfo  {publisher} {Springer},\ \bibinfo {year} {1998})\
  pp.\ \bibinfo {pages} {215--252}\BibitemShut {NoStop}%
\bibitem [{\citenamefont {Alt}\ \emph {et~al.}(2021{\natexlab{a}})\citenamefont
  {Alt}, \citenamefont {Ducatez},\ and\ \citenamefont
  {Knowles}}]{Alt_Ducatez_Knowles_2021}%
  \BibitemOpen
  \bibfield  {author} {\bibinfo {author} {\bibfnamefont {J.}~\bibnamefont
  {Alt}}, \bibinfo {author} {\bibfnamefont {R.}~\bibnamefont {Ducatez}}, \ and\
  \bibinfo {author} {\bibfnamefont {A.}~\bibnamefont {Knowles}},\ }\href
  {\doibase 10.1007/s00220-021-04167-y} {\bibfield  {journal} {\bibinfo
  {journal} {Communications in Mathematical Physics}\ }\textbf {\bibinfo
  {volume} {388}},\ \bibinfo {pages} {507–579} (\bibinfo {year}
  {2021}{\natexlab{a}})}\BibitemShut {NoStop}%
\bibitem [{\citenamefont {Alt}\ \emph {et~al.}(2021{\natexlab{b}})\citenamefont
  {Alt}, \citenamefont {Ducatez},\ and\ \citenamefont
  {Knowles}}]{Alt_Knowles_2021}%
  \BibitemOpen
  \bibfield  {author} {\bibinfo {author} {\bibfnamefont {J.}~\bibnamefont
  {Alt}}, \bibinfo {author} {\bibfnamefont {R.}~\bibnamefont {Ducatez}}, \ and\
  \bibinfo {author} {\bibfnamefont {A.}~\bibnamefont {Knowles}},\ }\href
  {http://arxiv.org/abs/2109.03227} {\  (\bibinfo {year}
  {2021}{\natexlab{b}})},\ \bibinfo {note} {arXiv:2109.03227
  [math-ph]}\BibitemShut {NoStop}%
\bibitem [{\citenamefont {Skraba}\ \emph {et~al.}(2020)\citenamefont {Skraba},
  \citenamefont {Thoppe},\ and\ \citenamefont
  {Yogeshwaran}}]{Skraba_Thoppe_Yogeshwaran_2020}%
  \BibitemOpen
  \bibfield  {author} {\bibinfo {author} {\bibfnamefont {P.}~\bibnamefont
  {Skraba}}, \bibinfo {author} {\bibfnamefont {G.}~\bibnamefont {Thoppe}}, \
  and\ \bibinfo {author} {\bibfnamefont {D.}~\bibnamefont {Yogeshwaran}},\
  }\href {http://arxiv.org/abs/1701.00239} {\  (\bibinfo {year} {2020})},\
  \bibinfo {note} {arXiv:1701.00239 [math]}\BibitemShut {NoStop}%
\bibitem [{\citenamefont {Myers}\ \emph {et~al.}(2019)\citenamefont {Myers},
  \citenamefont {Munch},\ and\ \citenamefont
  {Khasawneh}}]{myers2019persistent}%
  \BibitemOpen
  \bibfield  {author} {\bibinfo {author} {\bibfnamefont {A.}~\bibnamefont
  {Myers}}, \bibinfo {author} {\bibfnamefont {E.}~\bibnamefont {Munch}}, \ and\
  \bibinfo {author} {\bibfnamefont {F.~A.}\ \bibnamefont {Khasawneh}},\
  }\href@noop {} {\bibfield  {journal} {\bibinfo  {journal} {Physical Review
  E}\ }\textbf {\bibinfo {volume} {100}},\ \bibinfo {pages} {022314} (\bibinfo
  {year} {2019})}\BibitemShut {NoStop}%
\bibitem [{\citenamefont {Tran}\ \emph {et~al.}(2021)\citenamefont {Tran},
  \citenamefont {Chen},\ and\ \citenamefont
  {Hasegawa}}]{Tran_Chen_Hasegawa_2021}%
  \BibitemOpen
  \bibfield  {author} {\bibinfo {author} {\bibfnamefont {Q.~H.}\ \bibnamefont
  {Tran}}, \bibinfo {author} {\bibfnamefont {M.}~\bibnamefont {Chen}}, \ and\
  \bibinfo {author} {\bibfnamefont {Y.}~\bibnamefont {Hasegawa}},\ }\href
  {\doibase 10.1103/PhysRevE.103.052127} {\bibfield  {journal} {\bibinfo
  {journal} {Physical Review E}\ }\textbf {\bibinfo {volume} {103}},\ \bibinfo
  {pages} {052127} (\bibinfo {year} {2021})}\BibitemShut {NoStop}%
\bibitem [{\citenamefont {Saucan}(2020)}]{saucan2020discrete}%
  \BibitemOpen
  \bibfield  {author} {\bibinfo {author} {\bibfnamefont {E.}~\bibnamefont
  {Saucan}},\ }\href@noop {} {\bibfield  {journal} {\bibinfo  {journal} {arXiv
  preprint arXiv:2003.03844}\ } (\bibinfo {year} {2020})}\BibitemShut {NoStop}%
\bibitem [{\citenamefont {De~Silva}\ and\ \citenamefont
  {Carlsson}(2004)}]{de2004topological}%
  \BibitemOpen
  \bibfield  {author} {\bibinfo {author} {\bibfnamefont {V.}~\bibnamefont
  {De~Silva}}\ and\ \bibinfo {author} {\bibfnamefont {G.~E.}\ \bibnamefont
  {Carlsson}},\ }in\ \href@noop {} {\emph {\bibinfo {booktitle} {PBG}}}\
  (\bibinfo {year} {2004})\ pp.\ \bibinfo {pages} {157--166}\BibitemShut
  {NoStop}%
\bibitem [{\citenamefont {Shinaoka}\ \emph {et~al.}(2020)\citenamefont
  {Shinaoka}, \citenamefont {Geffroy}, \citenamefont {Wallerberger},
  \citenamefont {Otsuki}, \citenamefont {Yoshimi}, \citenamefont {Gull},\ and\
  \citenamefont {Kune{\v{s}}}}]{shinaoka2020sparse}%
  \BibitemOpen
  \bibfield  {author} {\bibinfo {author} {\bibfnamefont {H.}~\bibnamefont
  {Shinaoka}}, \bibinfo {author} {\bibfnamefont {D.}~\bibnamefont {Geffroy}},
  \bibinfo {author} {\bibfnamefont {M.}~\bibnamefont {Wallerberger}}, \bibinfo
  {author} {\bibfnamefont {J.}~\bibnamefont {Otsuki}}, \bibinfo {author}
  {\bibfnamefont {K.}~\bibnamefont {Yoshimi}}, \bibinfo {author} {\bibfnamefont
  {E.}~\bibnamefont {Gull}}, \ and\ \bibinfo {author} {\bibfnamefont
  {J.}~\bibnamefont {Kune{\v{s}}}},\ }\href@noop {} {\bibfield  {journal}
  {\bibinfo  {journal} {SciPost Physics}\ }\textbf {\bibinfo {volume} {8}},\
  \bibinfo {pages} {012} (\bibinfo {year} {2020})}\BibitemShut {NoStop}%
\end{thebibliography}%
\end{document}